\input harvmac

\input epsf
\overfullrule=0pt  
\parindent=0pt

\ifx\epsfbox\UnDeFiNeD\message{(NO epsf.tex, FIGURES WILL BE IGNORED)}
\def\figin#1{\vskip2in}
\else\message{(FIGURES WILL BE INCLUDED)}\def\figin#1{#1}\fi

%
%
%
\def\lb       {\left( }
\def\rb       {\right) }
\def\lmb      {\left\{ }
\def\rmb      {\right\} }
\def\lbb     {\left[ }
\def\rbb      {\right] }
\def\comma      { \, , }
\def\period     { \, . }
\def\bra#1      { \langle \, #1 \, \vert \, }
\def\ket#1      { \, \vert \, #1 \, \rangle  }
\def\semiket#1  { \, #1 \, \rangle \, }
\def\kket#1     { \, \vert \, #1 \, \rangle  \rangle }
\def\bbra#1       { \langle \! \langle \, #1 \, \vert \, }
\def\del        {  \partial  }
\def\delbar     { \bar{\partial} }
\def\zbar       { \bar{z} }
%
%
\def\ifig#1#2#3{\xdef#1{fig.~\the\figno}
\goodbreak\figin{\centerline{#3}}%
\smallskip\centerline{\vbox{\baselineskip12pt
\advance\hsize by -1truein\noindent{\bf
Fig.~\the\figno:} #2}}
\bigskip\global\advance\figno by1}

\newwrite\ffile\global\newcount\figno \global\figno=1
\def\fig{fig.~\the\figno\nfig}
\def\nfig#1{\xdef#1{fig.~\the\figno}%
\writedef{#1\leftbracket fig.\noexpand~\the\figno}%
\ifnum\figno=1\immediate\openout\ffile=figs.tmp\fi\chardef\wfile=
\ffile%
\immediate\write\ffile{\noexpand\medskip\noexpand\item{Fig.\
\the\figno. }
\reflabeL{#1\hskip.55in}\pctsign}\global\advance\figno by1\findarg}


%
%
\message{S-Tables Macro v1.0, ACS, TAMU (RANHELP@VENUS.TAMU.EDU)}
%
%
\newhelp\stablestylehelp{You must choose a style between 0 and 3.}%
\newhelp\stablelinehelp{You should not use special hrules when stretching
a table.}%
\newhelp\stablesmultiplehelp{You have tried to place an S-Table inside
another
S-Table.  I would recommend not going on.}%
%
%
\newdimen\stablesthinline
\stablesthinline=0.4pt
\newdimen\stablesthickline
\stablesthickline=1pt
%
%
\newif\ifstablesborderthin
\stablesborderthinfalse
\newif\ifstablesinternalthin
\stablesinternalthintrue
\newif\ifstablesomit
\newif\ifstablemode
\newif\ifstablesright
\stablesrightfalse
%
%
\newdimen\stablesbaselineskip
\newdimen\stableslineskip
\newdimen\stableslineskiplimit
%
%
\newcount\stablesmode
\newcount\stableslines
\newcount\stablestemp
\stablestemp=3
\newcount\stablescount
\stablescount=0
\newcount\stableslinet
\stableslinet=0
%
%
%
\newcount\stablestyle
\stablestyle=0
%
%
\def\stablesleft{\quad\hfil}%
\def\stablesright{\hfil\quad}%
%
%
\catcode`\|=\active%
%
%
\newcount\stablestrutsize
\newbox\stablestrutbox
\setbox\stablestrutbox=\hbox{\vrule height10pt depth5pt width0pt}
\def\stablestrut{\relax\ifmmode%
                         \copy\stablestrutbox%
                       \else%
                         \unhcopy\stablestrutbox%
                       \fi}%
%
%
\newdimen\stablesborderwidth
\newdimen\stablesinternalwidth
\newdimen\stablesdummy
\newcount\stablesdummyc
\newif\ifstablesin
\stablesinfalse
%
%
\def\begintable{\stablestart%
  \stablemodetrue%
  \stablesadj%
  \halign%
  \stablesdef}%
\def\stablesadj{%
  \ifcase\stablestyle%
    \hbox to \hsize\bgroup\hss\vbox\bgroup%
  \or%
    \hbox to \hsize\bgroup\vbox\bgroup%
  \or%
    \hbox to \hsize\bgroup\hss\vbox\bgroup%
  \or%
    \hbox\bgroup\vbox\bgroup%
  \else%
    \errhelp=\stablestylehelp%
    \errmessage{Invalid style selected, using default}%
    \hbox to \hsize\bgroup\hss\vbox\bgroup%
  \fi}%
\def\stablesend{\egroup%
  \ifcase\stablestyle%
    \hss\egroup%
  \or%
    \hss\egroup%
  \or%
    \egroup%
  \or%
    \egroup%
  \else%
    \hss\egroup%
  \fi}%
\def\stablestart{%
  \ifstablesin%
    \errhelp=\stablesmultiplehelp%
    \errmessage{An S-Table cannot be placed within an S-Table!}%
  \fi
  \global\stablesintrue%
  \global\advance\stablescount by 1%
  \message{<S-Tables Generating Table \number\stablescount}%
  \begingroup%
  \stablestrutsize=\ht\stablestrutbox%
  \advance\stablestrutsize by \dp\stablestrutbox%
  \ifstablesborderthin%
    \stablesborderwidth=\stablesthinline%
  \else%
    \stablesborderwidth=\stablesthickline%
  \fi%
  \ifstablesinternalthin%
    \stablesinternalwidth=\stablesthinline%
  \else%
    \stablesinternalwidth=\stablesthickline%
  \fi%
  \tabskip=0pt%
  \stablesbaselineskip=\baselineskip%
  \stableslineskip=\lineskip%
  \stableslineskiplimit=\lineskiplimit%
  \offinterlineskip%
  \def\borderrule{\vrule width \stablesborderwidth}%
  \def\internalrule{\vrule width \stablesinternalwidth}%
  \def\thinline{\noalign{\hrule height \stablesthinline}}%
  \def\thickline{\noalign{\hrule height \stablesthickline}}%
  \def\trule{\omit\leaders\hrule height \stablesthinline\hfill}%
  \def\ttrule{\omit\leaders\hrule height \stablesthickline\hfill}%
  \def\tttrule##1{\omit\leaders\hrule height ##1\hfill}%
  \def\stablesel{&\omit\global\stablesmode=0%
    \global\advance\stableslines by 1\borderrule\hfil\cr}%
  \def\el{\stablesel&}%
  \def\elt{\stablesel\thinline&}%
  \def\eltt{\stablesel\thickline&}%
  \def\elttt##1{\stablesel\noalign{\hrule height ##1}&}%
  \def\elspec{&\omit\hfil\borderrule\cr\omit\borderrule&%
              \ifstablemode%
              \else%
                \errhelp=\stablelinehelp%
                \errmessage{Special ruling will not display properly}%
              \fi}%
  \def\stmultispan##1{\mscount=##1 \loop\ifnum\mscount>3 \stspan\repeat}%
  \def\stspan{\span\omit \advance\mscount by -1}%
  \def\multicolumn##1{\omit\multiply\stablestemp by ##1%
     \stmultispan{\stablestemp}%
     \advance\stablesmode by ##1%
     \advance\stablesmode by -1%
     \stablestemp=3}%
  \def\multirow##1{\stablesdummyc=##1\parindent=0pt\setbox0\hbox\bgroup%
    \aftergroup\emultirow\let\temp=}
  \def\emultirow{\setbox1\vbox to\stablesdummyc\stablestrutsize%
    {\hsize\wd0\vfil\box0\vfil}%
    \ht1=\ht\stablestrutbox%
    \dp1=\dp\stablestrutbox%
    \box1}%
  \def\stpar##1{\vtop\bgroup\hsize ##1%
     \baselineskip=\stablesbaselineskip%
     \lineskip=\stableslineskip%

\lineskiplimit=\stableslineskiplimit\bgroup\aftergroup\estpar\let\temp=}%
  \def\estpar{\vskip 6pt\egroup}%
  \def\stparrow##1##2{\stablesdummy=##2%
     \setbox0=\vtop to ##1\stablestrutsize\bgroup%
     \hsize\stablesdummy%
     \baselineskip=\stablesbaselineskip%
     \lineskip=\stableslineskip%
     \lineskiplimit=\stableslineskiplimit%
     \bgroup\vfil\aftergroup\estparrow%
     \let\temp=}%
  \def\estparrow{\vfil\egroup%
     \ht0=\ht\stablestrutbox%
     \dp0=\dp\stablestrutbox%
     \wd0=\stablesdummy%
     \box0}%
  \def|{\global\advance\stablesmode by 1&&&}%
  \def\|{\global\advance\stablesmode by 1&\omit\vrule width 0pt%
         \hfil&&}%
  \def\vt{\global\advance\stablesmode by 1&\omit\vrule width
\stablesthinline%
          \hfil&&}%
  \def\vtt{\global\advance\stablesmode by 1&\omit\vrule width
\stablesthickline%
          \hfil&&}%
  \def\vttt##1{\global\advance\stablesmode by 1&\omit\vrule width ##1%
          \hfil&&}%
  \def\vtr{\global\advance\stablesmode by 1&\omit\hfil\vrule width%
           \stablesthinline&&}%
  \def\vttr{\global\advance\stablesmode by 1&\omit\hfil\vrule width%
            \stablesthickline&&}%
  \def\vtttr##1{\global\advance\stablesmode by 1&\omit\hfil\vrule width
##1&&}%
  \stableslines=0%
  \stablesomitfalse}
\def\stablesdef{\bgroup\stablestrut\borderrule##\tabskip=0pt plus 1fil%
  &\stablesleft##\stablesright%
  &##\ifstablesright\hfill\fi\internalrule\ifstablesright\else\hfill\fi%
  \tabskip 0pt&&##\hfil\tabskip=0pt plus 1fil%
  &\stablesleft##\stablesright%
  &##\ifstablesright\hfill\fi\internalrule\ifstablesright\else\hfill\fi%
  \tabskip=0pt\cr%
  \ifstablesborderthin%
    \thinline%
  \else%
    \thickline%
  \fi&%
}%
\def\endtable{\advance\stableslines by 1\advance\stablesmode by 1%
   \message{- Rows: \number\stableslines, Columns:  \number\stablesmode>}%
   \stablesel%
   \ifstablesborderthin%
     \thinline%
   \else%
     \thickline%
   \fi%
   \egroup\stablesend%
\endgroup%
\global\stablesinfalse}
%
%


\def\xxx#1 {{hep-th/#1}}
\def\lr { \lref}
\def\npb#1(#2)#3 { Nucl. Phys. {\bf B#1} (#2) #3 }
\def\rep#1(#2)#3 { Phys. Rept.{\bf #1} (#2) #3 }
\def\plb#1(#2)#3{Phys. Lett. {\bf #1B} (#2) #3}
\def\prl#1(#2)#3{Phys. Rev. Lett.{\bf #1} (#2) #3}
\def\physrev#1(#2)#3{Phys. Rev. {\bf D#1} (#2) #3}
\def\ap#1(#2)#3{Ann. Phys. {\bf #1} (#2) #3}
\def\rmp#1(#2)#3{Rev. Mod. Phys. {\bf #1} (#2) #3}
\def\cmp#1(#2)#3{Comm. Math. Phys. {\bf #1} (#2) #3}
\def\mpl#1(#2)#3{Mod. Phys. Lett. {\bf #1} (#2) #3}
\def\ijmp#1(#2)#3{Int. J. Mod. Phys. {\bf A#1} (#2) #3}
\def\mpla#1(#2)#3{Mod. Phys. Lett. {\bf A#1} (#2) #3}
\def\jhep#1(#2)#3{JHEP {\bf  #1} (#2) #3}

\parindent 25pt
\overfullrule=0pt
\tolerance=10000

\sequentialequations

\noblackbox
\baselineskip 14pt plus 2pt minus 2pt
\Title{\vbox{\baselineskip12pt
\hbox{hep-th/9808080}
\hbox{PUPT-1808}
}}
{\vbox{
\centerline{D-branes in Gepner models and supersymmetry} }}

\vskip 5ex
\centerline{Michael Gutperle\foot{email: gutperle@feynman.princeton.edu}
and Yuji Satoh\foot{email: ysatoh@viper.princeton.edu }}
\medskip
\centerline{Department of Physics, Princeton University, Princeton NJ
08554, USA} \bigskip

\vskip 15ex
\medskip
\centerline{{\bf Abstract}}
\vskip 5ex
Boundary states corresponding to wrapped D-branes in Calabi-Yau
compactifications of type II strings are discussed using  Gepner models.
In particular  boundary conditions corresponding  to D-0 branes and
D-instantons in four dimensions are investigated. 
The boundary states constructed by
Recknagel and Schomerus are analyzed in the light-cone gauge and the broken
and conserved space-time supersymmetry charges are found. 
The geometrical interpretation of these algebraically constructed
boundary states is clarified in some simple cases. 
 Moreover, the action of  
mirror symmetry and other discrete symmetries of the Gepner model on the
boundary states are discussed. As an application the boundary states are
used to calculate instanton induced corrections to metric on the
hypermultiplets in the $N=2$ effective action.

\noblackbox
\baselineskip 14pt plus 2pt minus 2pt

\Date{August 1998}
\vfill\eject

\lr\banks{T. Banks and L.J. Dixon, {\it Constraints on string vacua with
    space-time supersymmetry}, \npb307(1988)93.}
\lr\scft{W. Boucher, D. Friedan and A. Kent, 
   {\it Determinant formula and unitarity for the N=2 superconformal 
    algebras in two-dimensions or exact results on string compactification}, 
    \plb172(1986)316; 
         ~S. Nam, {\it The Kac formula for the N=1 and the N=2 
                 superconformal algebras}, \plb172(1986)323; 
         ~P. di Vecchia, J.L. Petersen, M. Yu and H.B. Zheng, 
        {\it Explicit construction of unitary representations of the N=2 
             superconformal algebra}, \plb174(1986)280.}
\lr\spectralfl{A. Schwimmer and N. Seiberg, {\it Comments on the
    $N=2,N=3,N=4$ superconformal algebras in two dimensions},
\plb184(1987)191.}
\lr\lerche{W. Lerche, C. Vafa, N.P. Warner, {\it Chiral rings in $N=2$
    superconformal theories}, \npb324(1989)427.}
\lr\gepner{D. Gepner, {\it Space-time supersymmetry in compactified
    string theory and superconformal models}, \npb296(1988)757; {\it
    Exactly solvable string compactifications on manifolds of $SU(N)$
    holonomy}, \plb199(1987)380.}
\lr\reck{A. Recknagel and  V. Schomerus, {\it D-branes in Gepner models},
\xxx9712186.} 
\lr\cardy{J.L. Cardy, {\it Boundary conditions, fusion rules and the
Verlinde formula}, \npb324(1989)581.}
\lr\greenepless{B.R. Greene and  M.R. Plesser, {\it Duality in Calabi-Yau
moduli spaces}, \npb338(1990)15.}
\lr\candelas{P. Candelas, M. Lynker and R. Schimmrigk, 
   {\it Calabi-Yau manifolds in weighted $P(4)$}, \npb341(1990)383.}
\lr\greenlec{B.R. Greene, {\it String theory on Calabi-Yau manifolds},
\xxx9702155.}
\lr\greengut{M.B. Green and  M. Gutperle, {\it Light cone supersymmetry
and D-branes}, \npb476(1996)484, \xxx9604091.}
\lr\ademodels{J. Fuchs, A. Klemm, C.  Scheich and  M.G. Schmidt, 
{\it Spectra and symmetries of Gepner models compared to
    Calabi-Yau compactifications}, \ap204(1990)1.} 
\lr\polchinski{J. Polchinski, 
   {\it Dirichlet branes and Ramond-Ramond charges}, 
   \prl75(1995)4724, \xxx9510017.}
\lr\boundstform{C.G. Callan, C. Lovelace, C.R. Nappi and S.A. Yost, 
       {\it Adding holes and crosscaps to the superstrings}, 
        \npb293(1987)83;  
        ~J. Polchinski and Y. Cai, 
        {\it Consistency of open superstring theories}, \npb296(1988)91.}
\lr\ishibashi{N. Ishibashi, {\it The boundary and crosscap states in
    conformal field theories}, \mpla4(1989)251.}
\lr\wittentop{E. Witten, {\it Mirror manifolds and topological field
    theory}, \xxx9112056.}
\lr\conifold{A. Strominger, 
{\it Massless black holes and conifolds in string theory}, \npb451(1995)96.}
\lr\greengutb{M.B.~Green and M.~Gutperle, {\it Effects of D-instantons}, 
 \npb498(1997)195, \xxx9701093.}
\lr\wittenb{E. Witten, {\it Phases of N=2 theories in two-dimensions},
    \npb403(1993)159.}
\lr\chiral{E. Witten, {\it Dynamical breaking of supersymmetry}, 
    \npb188(1981)513.} 
\lr\yukawa{D. Gepner, {\it Yukawa couplings for Calabi-Yau string
    compactification}, \npb311(1988)191; J. Distler and  B. Greene {\it
    Aspects of $(2,0)$ string compactifications}, \npb304(1988)1.}
\lr\cecotti{S. Cecotti, S. Ferrara and  L. Girardello, {\it Geometry of
type-II superstrings and the moduli space of superconformal field
theories},  \ijmp4(1989)2475. }
\lr\lust{K.Behrndt et al., {\it From type IIA black holes to T-dual
    type IIB D instantons in N=2, D = 4 supergravity}, \npb508(1997)659, 
   \xxx9706096. }
\lr\greenevafa{B.R. Greene, C. Vafa, N.P. Warner, {\it Calabi-Yau
    manifolds and renormalization group flows}, \npb324(1989)371.}
\lr\bbs{K.~Becker, M.~Becker and  A.~Strominger, {\it Fivebranes,
 membranes and non-perturbative string theory}, 
  \npb456(195)130, \xxx9507158. }
\lr\oogurioz{ H.~Ooguri, Y.~Oz and Z.~Yin, {\it D-Branes on Calabi-Yau
 spaces and their mirrors}, \npb477(1996)407,  \xxx9606112. }
\lr\sagnotti{G. Pradisi, A. Sagnotti and  Ya.S. Stanev, {\it
    Completeness conditions for boundary operators in 2-D conformal
    field theory}, \plb381(1996)97, \xxx9603097. }
\lr\typeI{C. Angelantonj et al,
    {\it  Comments on Gepner models and type I vacua in string
    theory}, \plb 387(1996)743; R. Blumenhagen and A. Wisskirchen, 
    {\it Spectra of 4D, N=1 type I string vacua on non-toroidal 
     CY threefolds}, \xxx9806131.}
\lr\gepnerlec{D. Gepner, {\it Lectures on $N=2$ string theory}, in
Superstrings '89, ed. M.B. Green et al, World Scientific, p.238-302.}
\lr\oogurivafa{H.~Ooguri and C.~Vafa, {\it Summing up D-instantons},  
\prl77(1996)3296, \xxx9608079;  ~B.R.~Greene, D.R.~Morrison and C.~Vafa,
{\it A geometric realization 
of confinement}, \npb481(1996)513, \xxx9608039.}
\lr\wittenbc{E.~Witten, {\it  Chern-Simons gauge theory as a string theory}, 
 \xxx9207094}
\lr\polchdint{J. Polchinski, {\it  Combinatorics of boundaries in string
theory}, \physrev50(1994)6041, \xxx9407031.}
\lr\mbggas{M.B. Green {\it  A gas of D-instantons}, \plb354(1995)271,
\xxx9504108. }
\lr\greengutd{M.B. Green and M. Gutperle, {\it  D instanton partition
functions}, \xxx9804123.}
\lr\anton{I. Antoniadis, B. Pioline and T.R. Taylor, {\it Calculable
$e^{-1/\lambda}$ effects} , \npb512(1998)61, 
\xxx9707222.}
\lr\ttwm{ T. Eguchi and S.-K. Yang, 
   {\it N=2 superconformal models as topological field theories}, 
    \mpla5(1990)1693.}
\lr\GS{A. Giveon and D-J. Smit,
             {\it Symmetries on the moduli space of (2,2) superstring vacua}, 
             Nucl. Phys. {\bf B349} (1991) 168.}
\lr\Warner{N.P. Warner, 
         {\it Lectures on N=2 superconformal theories and singularity theory},
          the book in \gepnerlec, p.197-237.}
\lr\Qiu{Z. Qui, {\it Nonlocal current algebra and N=2 
                        superconformal field theory},
           Phys. Lett. {\bf B188} (1986) 207.}
\lr\fuchs{J. Fuchs and C. Schweigert, {\it Branes: From free fields to general backgrounds}, \xxx 9712257.}
\newsec{Introduction}

The compactification of type II string theories   
  on a Calabi-Yau
threefold  produces four dimensional  $N=2$ supersymmetric 
 theories.  Many
 interesting phenomena like mirror symmetry, special geometry and string
 duality arise in these theories.
Dirichlet branes \polchinski\  provide a simple realization of
non-perturbative
solitonic objects carrying  Ramond-Ramond
($R\otimes R$) charges in type II and type I string theories.  The
description of 
D-branes wrapping  submanifolds of Calabi-Yau manifolds is 
an important ingredient for  non-perturbative string theory. One example is
Strominger's  resolution of the conifold singularity \conifold.
Euclidean wrappings  of branes on cycles in  Calabi-Yau manifolds 
were discussed by Becker, Becker and
Strominger  \bbs.  Using the world-volume description of
branes geometric  criteria for the wrapping  cycles to
preserve half the supersymmetry were found.   The wrapping of D-branes on
such `supersymmetric cycles' 
 was analyzed further 
by  Ooguri, Oz and Yin \oogurioz. In these references the  Calabi Yau
compactifications are
discussed in a sigma model framework where only the long wavelength
(massless) excitations are taken into account. 

On the other hand   an
exactly solvable model  of a
 compactification on  a complex manifold of dimension $2k,k=1,2,3$  with 
holonomy group $SU(k)$ 
was constructed by  Gepner \gepner.  In this construction the 
internal  $c=3k$  $N=2$ superconformal field theory (SCFT) is made of a
tensor
product  of $N=2$ minimal models with the correct total central charge. 
Since the Gepner model is exactly solvable many techniques for
constructing boundary states  in rational conformal field theories
\ishibashi,\cardy,\sagnotti\  can be applied in this
context.\foot{
Gepner models in type I theory context were discussed in \typeI.}
In particular Recknagel and Schomerus \reck\ constructed 
boundary states for  Gepner models. With these
boundary states  
it may be possible to discuss many interesting
problems in a precise manner including the massive string excitations.
In addition, D-branes have played an important role
in recent development in black hole physics. 
In this respect, the boundary states
in Gepner models may give useful insight for understanding black holes 
in string theory in Calabi-Yau compactifications.
 In this paper we will discuss only trivial, 
i.e. $U(1)$, Chan-Paton factors which corresponds 
to the wrapping of a single D-brane.

However, the physical aspects of the boundary states 
in the string theory context have not 
been fully explored yet. In this paper, we will address this issue
and analyze the boundary states of \reck\ further  
from a space-time perspective. 
In particular we will use  space-time supersymmetry
in the light-cone gauge \greengut\  to construct the broken and unbroken
supersymmetries associated with the wrapped brane. Furthermore the
geometric interpretation of the boundary states in Gepner models will be
discussed. We will find what D-branes those algebraically constructed boundary 
states represent in some simple cases.  
Gepner models have interesting discrete
symmetries which are inherited from the symmetries of the minimal models. One important example is mirror symmetry. We will 
discuss how these symmetries are realized in the boundary states.
We will also find some non-perturbative effects induced by D-instantons.

The organization of the paper is as follows.
In section 2  we will give a brief review of $N=2$ space-time
supersymmetry  and its relation to the internal (2,2) superconformal
symmetry. In section 3 the Gepner model
construction is reviewed, see \gepnerlec\ for more details.  In section 4
the
boundary states are presented. In section 5 the case of  the Gepner
models corresponding to compactifications on $T^2$ is discussed and the
geometric interpretation of the boundary states is given. 
 Broken and unbroken supersymmetries are analyzed and a criterion for
 mutually supersymmetric  D-brane configuration is found in section
 6. In section 7 the action of mirror symmetry and T-duality
 (c-map) on the boundary state is discussed.  The more complicated
 case of the $(k=3)^5$ Gepner model which corresponds to a special
 quintic hypersurface is discussed in  section 8. 
In section 9  some non-perturbative effects
 induced by D-instantons  are calculated using the boundary states and
 the broken  supersymmetry charges. 
The last section contains some comments
 about issues not discussed in the main part of the paper and
 conclusions. In appendix A, 
we give some calculation of the partition functions 
from the boundary states for simple Gepner models.
In appendix B, we discuss geometrical interpretation of 
the boundary conditions for the space-time fields by using the free field 
realizations of $N=2$ minimal models.

\newsec{$N=2$ compactification and space time supersymmetry}

An $N=2$ super conformal algebra  with central charge $c$ contains
apart from the stress tensor $T$ also  two supercurrents $G^\pm$
 and a 
 $U(1)$ current $J$ \scft. 
The current $J$  can be expressed in
term of a free bosonic current in the following way 
\eqn\jone{J= i\sqrt{c\over 3}\partial H \period } 
The $N=2$ algebra  admits an automorphism called spectral
 flow  \spectralfl, which is given by twisting with respect to the
 $U(1)$  current. Spectral flow by half a unit connects  the
 Neveu-Schwarz (NS) and the Ramond (R) sectors of the theory.
The generators of the spectral flow are 
\eqn\specfl{U_{+1/2}= \exp(+{i\over 2}\sqrt{c\over 3}H),\quad
 U_{-1/2}= \exp(-{i\over 2}\sqrt{c\over 3}H) \period }
A field $\Phi$ of the $N=2$ SCFT has a specific conformal dimensions $h$
and
 $U(1)$ charge $q$. The properties of the $N=2$ algebra imply that
 there is a lower bound for the conformal dimension  $h\geq
 1/2 \mid q\mid $ for every field $\Phi$. Fields  which saturate this
bound $h=1/2 \mid q \mid
 $ are primary. The field is  annihilated by either $G_{-1/2}^+$ or
 $G_{-1/2}^-$ and  called chiral (c) or antichiral (a) respectively.
 (Anti)chiral primaries have nonsingular OPE's among themselves and
 form  a finite ring, called (anti)chiral ring  \lerche. 

 An
internal conformal field theory with central charge  
$c_{int}=12-3(D-2)/2$ is
 used to compactify type II superstring theory to $D$ non-compact 
dimensions. The compactified theory will be  space-time supersymmetric
only if  the internal
 conformal field theory has (at least)  $N=2$ superconformal
invariance \banks. 

Hence compactification to four dimensions ($D=4$) gives an 
 internal  SCFT with $c=9$ and a transverse SCFT with $c=3$. 
The transverse SCFT  describes
 the  propagation of the string in four dimensional Minkowski
 space in  the  light-cone gauge, where  
 all string excitations  can be described
 in terms of 
 a free complex boson $X=X^1+iX^2$ and fermion $\psi=\psi^1+i\psi^2$. This
 system is the simplest  example of an $N=2$ SCFT with $c=3$ where
 the superconformal  
tensors are given by
\eqn\ceqthree{T= -{1\over 2}\partial X\partial X^*+{1\over 2} \psi
\partial \psi^*+ {1\over 2} \psi^* \partial \psi,
\quad G^+= \psi\partial X^*,\quad G^-= \psi^*\partial X, \quad J_{ext}=
 \psi^*\psi \period } 
Using \jone\ the  $U(1)$ currents  $J_{ext}$ and $J_{int}$ for the
transverse ($c=3$) and
 internal ($c=9$) SCFT can be expressed in terms of a 
 free
 boson $\phi$ and    a  free
 boson $H$ respectively, 
\eqn\uonefree{J_{ext}=i\partial \phi, \quad J_{int}=i\sqrt{3}\partial H.}
The transverse $U(1)$ charge determines  the  
helicity of the transverse 
state in the light-cone gauge.
The operators implementing  the spectral flow by $\eta = \pm 1/2$ 
are given by  
 $SO(2)$ spin fields  which connect  the NS sector and  the R sector, 
\eqn\spinfield{ S= \exp( +i {1\over 2}\phi), \quad S^\dagger= \exp( -i
  {1\over 2}\phi) \period}
Space-time supersymmetry is achieved by imposing a generalized GSO
projection
which keeps only   states with  odd integer charge
with respect to the total $U(1)$ current  given by the sum of the
internal and space-time \uonefree\ currents, 
\eqn\totuone{J_{tot}= J_{ext}+J_{int}=i \partial 
\phi+ i \sqrt{3}\partial H \period }
 In
the  light-cone gauge the four supersymmetry charges $Q^a, a=1,\cdots,4$
can be divided  into linearly realized ones for which $\Gamma^+
Q=0$  and nonlinearly realized ones for which $\Gamma^- Q=0$. The
linear  supercharges are constructed from the transverse spin fields  
\spinfield\ and the spectral flow operator \specfl\
\eqn\superch{\eqalign{Q&= \sqrt{p^+}\oint\exp( +i {1\over 2}\phi) \exp(i
  {\sqrt{3}\over 2}H) \comma \cr
  Q^\dagger&= \sqrt{p^+}\oint\exp( -i 
{1\over 2}\phi) \exp(-i{\sqrt{3}\over 2}H) \period }}
The nonlinear supercharges are given by
\eqn\supchnl{\eqalign{S&=  {1\over\sqrt{p^+}}\oint \exp( +i {1\over
 2}\phi)\exp(i{\sqrt{3}\over 2}H) \partial X \comma \cr
 S^\dagger&=   {1\over\sqrt{p^+}}\oint\exp( -i {1\over 2}\phi)
\exp(-i{\sqrt{3}\over 2}H) \partial X^* \period }}
The supercharges satisfy the anti-commutation relations of $N=1$
supersymmetry algebra  in light-cone coordinates,   
\eqn\comrel{\{ Q,Q^\dagger\}=p^+,\quad \{S,S^\dagger\}=p^-, \quad
\{Q^\dagger,S\}= p,\quad\{Q, S^\dagger\}=p^* \period } 
Where $p^-$ is determined by the physical state  conditions in terms of
the
zero mode of the energy momentum tensor and $p=p_1+ip_2,p^*=p_1-ip_2$
denotes the
 transverse momenta in a complex basis.

\bigskip
\begintable
\hbox{rep}|$\lambda$|$\Delta_{ext}$|$q_{int}$| $\Delta_{int}$\elt
o|0|0|$\pm 1$|1/2\elt
v|$\pm 1$|1/2|0|0\elt
s|+1/2|1/8|1/2,-3/2|3/8\elt
c|-1/2|1/8|-1/2,+3/2|3/8
\endtable
\centerline{{\bf Table 1}: Massless left-moving sectors for a $c=9$
  compactification. $\lambda$ and $\Delta_{ext}$ denote}
\centerline{ external charge and
  dimension whereas $q$ and $\Delta_{int}$ denote internal charge and
  dimension.}
\bigskip

The spectrum of type II string theories  is  a tensor product of 
 left-moving
and right-moving sectors. In the following we shall use unbarred fields for
the left-movers and barred fields for the right-movers. The massless states
in $N=2$ $D=4$
compactifications are the ones which have left and
the right  conformal dimension
$\Delta_{tot}= \Delta_{ext}+\Delta_{int}=1/2,
\bar{\Delta}_{tot}=\bar{\Delta}_{ext}+\bar{\Delta}_{int}=1/2$. Such states
are labeled by the left and right-moving
helicities $\lambda,\bar{\lambda}$ and $U(1)$ charges $q,\bar{q}$

 Type IIA
and type IIB strings differ by  the relative sign of the internal and
external $U(1)$ charge in the total $U(1)$ current for the
right-movers. The left-moving current is given by \totuone\ and the right
moving current for IIB and IIA is given by 
\eqn\totright{\eqalign{IIB:&\quad
   \bar{
J}_{tot}=i\bar{\partial}\bar{\phi}+i\sqrt{3}\bar{\partial}\bar{H} \comma 
 \cr
IIA:&\quad \bar{J}_{tot}=i\bar{\partial}\bar{\phi}
   -i\sqrt{3}\bar{\partial}\bar{H} \period }}
The difference between IIB and IIA is the reversal of the sign of
$\bar{H}$. This can be traced back to the fact that IIB in ten dimensions
is chiral whereas IIA is not.
The massless spectrum is  determined by tensoring the states
given in table 1 for left and right-movers and projecting onto odd integer 
$U(1)$ charges  using the currents \totuone\ and \totright.

 For type IIA, the right-moving supersymmetry charges are then similarly 
given by reversing the sign of $\bar{H}\to -\bar{H}$ in 
 \superch,\supchnl. This implies that   
 the supersymmetry  $Q$
maps $\lambda\to \lambda+1/2$ and $q \to q+3/2$ and $\bar{Q}$
maps $\bar{\lambda}\to \bar{\lambda}+1/2$ and $\bar{q}\to \bar{q}-3/2$. 
In contrast, for
IIB the left and right signs are the same and  $Q$
maps maps $\lambda\to \lambda+1/2$ and $q \to q+3/2$ and $\bar{Q}$
maps $\bar{\lambda}\to \bar{\lambda}+1/2$ and $\bar{q}\to \bar{q}+3/2$. 
The action of the supersymmetries relate the scalars in the $NS\otimes
NS$ sector to the $R\otimes R$ sector. $R\otimes R$ states with
$\lambda+\bar{\lambda}=0$ belong into  hypermultiplets whereas states
with $\lambda+\bar{\lambda}= \pm 1$ fall into  vector multiplets. Table
2 shows  which chiral rings give vector and  
hypermultiplets in the
massless sector. Note that the role of the chiral rings is interchanged
for IIA and IIB.

\bigskip
\begintable
\hbox{ type }|vector|hyper\elt
IIA|(a,c)+(c,a)|(c,c)+(a,a)\elt
IIB|(c,c)+(a,a)|(a,c)+(c,a)
\endtable
\medskip
\centerline{{\bf Table 2}: Hyper and vector multiplets for type IIA/B}
\bigskip
Using the sigma-model description of Calabi-Yau compactification the
elements of the (anti)chiral rings can be associated with elements of
the cohomology classes of the Calabi-Yau manifold and  it can be shown
 that the $(c,c)$ ring is in one to one correspondence with  
$H_{2,1}$ and the $(a,c)$ ring is in one to one correspondence with
$H_{1,1}$ \chiral.

\newsec{Review of Gepner models}
Gepner models \gepner\ are exactly soluble supersymmetric
compactifications of
type  II and heterotic strings which use tensor products of $N=2$
minimal  models to construct the internal SCFT. The $N=2$ minimal
models  are unitary representations of the $N=2$ SCFT which are
labeled  by an integer $k=1,2,\cdots$ where the central charge is given by
\eqn\centrch{c={3k\over k+2} \period }
Primary fields  $\Phi^l_{m,s}$ are labeled by three integers $l,m,s$
with the ranges\foot{Note that states with $s=2$ are really 
descendants. Nevertheless
  splitting each  module into subsets with $s=0$ and $s=2$ is a very
useful
  bookkeeping device. }
\eqn\range{l=0,1,\cdots,k,\quad  m=-(k+1),\cdots, k+2,\quad s=0,2,\pm 1 
\period }
together with  constraint $l+m+s\in 2Z$. The field 
identifications $(l,m,s)\sim (l,m,s+4)$
and $( l,m,s)\sim(l,m+2(k+2),s$) imply  that $m$ is defined modulo
$2(k+2)$ and $s$ is defined modulo 4. The labels $(l,m,s)$ can be
 brought into the
 `standard range' by another field identification
$(l,m,s)\sim (k-l,m+k+2,s+2)$.
The conformal dimension $h$  and $U(1)$ charge $q$
 of the primary fields (with $(l,m,s)$ in the standard rage) are given by
\eqn\confch{\eqalign{h&= {l(l+2)-m^2\over 4(k+2)}+{s^2\over 8} \comma \cr
q&={m\over k+2}-{s\over 2} \period }}
A Gepner model is constructed by tensoring $n$ minimal models with
$k_i, i=1,\cdots, n$ such that the sum  of the central charges of the
$n$ minimal models is equal to 
\eqn\ctot{\sum_{i=1}^n{3k_i\over k_i+2}= c_{int} \period }
The total currents $T,G^{\pm},J$ of the tensor product are given 
by the sum
of the currents of each minimal model.
The external theory is given by $D-2$ free bosons and a level one
$SO(D-2)$ current algebra. The primary fields can be labeled by two
vectors
\eqn\lamvec{\lambda = (l_1,\cdots,l_n),\quad
\mu=(s_0;m_1,\cdots, m_n;s_1,\cdots,s_n) \period}
Here $s_0=0,2,+1,-1$ labels the four characters corresponding to
$o,v,s,c$ conjugancy classes
 of the $SO(D-2)$ current algebra. Gepner constructed a  supersymmetric
 partition function for the tensor product  by using charge projections
 (generalizing the GSO projection) and adding
 twisted sectors to achieve  modular invariance. This  `$\beta$-method' uses
the
 $2n+1$ dimensional vectors:   $\beta_0$ which has $1$ everywhere  and
 $\beta_i,i=1,\cdots,n$ which has $2$ in the first and $n+1+j$ entry and 
is zero everywhere else. An inner product of two $2n+1$ dimensional
vectors is defined  by\foot{Here and in the following we display the formulae for $D=4$ and $D=8$. The construction for $D=6$ 
is slightly different and will not be needed  in this paper.} 
\eqn\inprod{\mu\bullet \tilde{\mu}= -{(D-2)\over 8}s_0\tilde{s}_0-
\sum_{j=1}^n {s_j\tilde{s_j}\over 4}+\sum_{j=1}^n {m_j\tilde{m}_j\over
2(k_j+2)}\period}
Note that with the help of this inner product the total $U(1)$ charge
of a  primary field is given by $q_{\mu}=2\beta_0\bullet \mu$. The GSO
projection is then implemented by projecting onto states with
an odd integer charge $q_{\mu}$. In order to preserve the
$N=1$  superconformal invariance all fields in the tensor product have
to be  in the same sector (R or NS). This can be achieved by
projecting onto  states which satisfy $\beta_j\bullet \mu\in Z$ for
$j=1,\cdots n$. Gepner constructed  a modular invariant partition
function by including twisted sectors,   
\eqn\partf{Z= {1\over 2^n} {({\rm Im } \ \tau)^{-(D-2)}\over
    \mid\eta(q)\mid^{2(D-2)}}\sum_{b_0,b_j}\sum^{\beta}_{\lambda,\mu}
  (-1)^{b_0}\chi^\lambda_{\mu}(q)\chi^{\lambda}_{\mu+b_0\beta_0+\sum_jb_j   
    \beta_j}(\bar{q}) \period }
Here $b_j=0,1$;  $b_0=0,1,\cdots, K-1$;  $K= $lcm$(4,2(k_j+2))$
and $ q = e^{2\pi i \tau}$. 
$ \chi_\mu^\lambda $ are the characters corresponding to the primaries
$\Phi_\mu^\lambda$. In
\partf\ the   diagonal affine  $SU(2)$ invariant is used which exists for
all levels  $k_j$. Other choices according to the ADE classification
of affine $SU(2)$ invariants 
 are possible and lead to different models \ademodels.
The notation $\sum^{\beta}$ indicates  the summation over the $\beta$
projected range ${\lambda,\mu}$ and the $(-1)^{b_0}$ imposes the
connection between spin and statistics.
Note that the supersymmetries \superch\ have a very simple action on
the characters $\chi^\lambda_\mu$; acting with $Q$ corresponds to
$\mu\to \mu+\beta_0$ and acting with $Q^\dagger$ corresponds to $\mu\to
\mu-\beta_0$.

Evidence for the equivalence of a Gepner model to compactification on
a  Calabi Yau manifold  was first presented in \gepner. 
  It was shown that 
massless
spectrum and the discrete symmetries of 
 Gepner models and certain hypersurfaces in weighted
projective spaces and  orbifolds thereof are the same.  Using \confch\ it
is easy
to 
see that chiral and antichiral primaries in the Gepner model 
correspond (up to field identification) to fields 
$\Phi^{\lambda,\lambda}_{\mu,{\mu}}$ and  
$\Phi^{\lambda,\lambda}_{\mu,-{\mu}}$ with
$l_i=m_i, s_i=0$  respectively. The massless fields satisfy
$2\beta_0\bullet \mu = \pm 1$. 
Furthermore the Yukawa couplings are the same \yukawa. The equivalence was
put onto firmer footing using a linear sigma model \wittenb\ 
 which interpolates
between the CY sigma model and the  LG orbifolds which in turn are
well known to be equivalent to minimal models \greenevafa.

In the case of $D=8$ $(c=3)$ only three Gepner models exist,
$(k=2)^2$, $(k=1)^3$  and $(k_1=1,k_2=4)$. The first one
corresponds to a compactifications on a $SU(2)^2$ torus and the last
two   correspond to a compactification on a $SU(3)$ torus.
For  $D=4$  $(c=9)$ a large (but still finite) number of models exist
corresponding  to Calabi-Yau compactifications. As an
specific example we will use the $(k=3)^5$ model which corresponds to
the quintic hypersurface in $CP^4$.

\newsec{Boundary states for  Gepner models}
Dirichlet branes \polchinski\ provide a surprisingly simple
  realization of  non-perturbative objects in
closed string theories.  D-branes  can be described  by the  
boundary state
  formalism \boundstform\  where   open string boundary conditions are
  enforced   on the
  closed string fields.
 Consistent boundary conditions for
superstrings require that the  $N=1$
superconformal invariance is  unbroken. This  implies continuity
conditions for the stress tensor and its superpartner  on the boundary. In
the
simplest  case of the upper half plane ${\cal H}=\{z\mid {\rm Im} (z)>0\}$, 
we require
$T(z)=\bar{T}(\bar{z})$  and $G(z)=\pm \bar{G}(\bar{z})$ at the
boundary  $z=\bar{z}$. Via a conformal  transformation which maps the
upper half plane into a semi-infinite cylinder this can be
related  to conditions on a boundary state $\mid B\rangle$.

If the conformal theory forms an extended algebras boundary conditions
relating the left and right-moving 
 currents in the extended algebra $W,\bar{W}$
have to be specified  \ishibashi. 
To construct  boundary states for rational conformal field
theories  one first defines Ishibashi states  $\mid i \rangle \rangle $ 
for every
primary field defining a irreducible highest weight
representation ${\cal H}_i$ of the algebra  
which satisfy \ishibashi\ 
\eqn\symalg{\big( W_n-(-1)^{h_W}\bar{W}_{-n}\big)\mid i\rangle \rangle = 0
\period } 
 In \ishibashi\  it was shown that  an Ishibashi
state can be constructed using an anti-unitary operator $U$ which 
acts on the
modes of the right-moving current $\bar{W}$ in the following way,
$U \bar{W}_n U^{-1}= (-1)^{h_W} \bar{W}_n$.
When $\bar{W} $ is a fermionic operator, we need to take into account 
the anti-commutativity to prove \symalg, and thus the action of $U$ needs
extra phase $(-1)^{F}$ where $F$ is the fermion-number operator.
Such an operator $U$ is closely related to the chiral CPT operator. 
Explicit form of the Ishibashi state is given by
\eqn\ishibashia{\mid i\rangle\rangle= \sum_N \mid i,N\rangle \otimes
U\widetilde{\mid i,N\rangle \period }}
where $N$ denotes the sum over the basis of ${\cal{H}}_i$.
In the second step a boundary state can be 
 constructed from a complete set of Ishibashi
states 
\eqn\bounstatea{\mid \alpha \rangle= \sum_i B^\alpha_i \mid i
  \rangle\rangle \period }
There are constraints on $B^\alpha_i$ which come from the fact that a
boundary state
$\mid \alpha \rangle$ has to define  an open string boundary condition.
This
implies that the modular transform of 
the cylinder amplitude $Z_{\alpha\beta}(q)=\langle \beta \mid
e^{-\pi t H_{cl}}\mid \alpha \rangle$ is related to an open string
partition function $Z_{\alpha\beta}(\tilde{q})=
Tr_{H_{\alpha\beta}}e^{-\pi \tau H_{op}}$ via a modular transformation
$t=1/ \tau$. The consistency of open
string partition function demands that it contains the characters of the
unbroken
symmetry algebra with integer multiplicities and this imposes nonlinear
constraints on the matrix $B^\alpha_i$. A solution to these constraints
was found by Cardy \cardy,
\eqn\cardsol{B^\alpha_i= {S_{\alpha i}\over \sqrt{S_{0i}} } \period }
Here $S_{ai}$ is the modular $S$-matrix and $0$ denotes the vacuum
representation.

In \oogurioz,\wittenbc\ it was shown that two different boundary
conditions for $U(1)$ current $J$ and the  superconformal generators
$G^\pm$  are
consistent with  $N=1$ superconformal invariance.
The two cases are called  A and B boundary conditions, referring to the two
possible topological twists of the $N=2$ theory \wittentop. The A
boundary conditions are defined by
\eqn\Atypebc{(J_n-\bar{J}_{-n})\mid B\rangle=0,\quad (G^-_r+i\eta
  \bar{G}^+_{-r})\mid B\rangle=0 \comma }
whereas the B type boundary conditions are  defined  by
 \eqn\Btypebc{(J_n+\bar{J}_{-n})\mid B\rangle=0,\quad  (G^+_r+i\eta
  \bar{G}^+_{-r})\mid B\rangle=0 \period }
The choice of
$\eta=\pm  1$ corresponds to a choice of spin structure. 
The anti-unitary operator $U$ in \ishibashia\  acts on  the operators of
the $N=2$ algebra in the following way 
\eqn\uaction{U^{-1} \bar{J}_n U= -  \bar{J}_n, \quad U^{-1} \bar{G}^\pm_r
U= -i \eta \bar{G}^{\mp}_r (-1)^{F} \period }
It is therefore easy to see that an  Ishibashi state for 
an $N=2$
minimal model  \ishibashia\ using $U$  imposes  $B$
boundary  conditions which satisfy  $q =- \bar{q}$.
On the other hand  it follows from \Atypebc\ that A  boundary conditions
satisfy $q = \bar{q}$.  Such boundary conditions are obtained from 
B boundary conditions
by a twist ${\Omega}$ which undoes the charge
reversal caused by $U$. This is given by the mirror automorphism
$\Omega$ of the $N=2$ algebra,
\eqn\mirraut{\Omega^{-1}\bar{J}_n\Omega=- \bar{J}_n,\quad
  \Omega^{-1}\bar{G}^\pm_r \Omega= \bar{G}^\mp_r \period }
An Ishibashi state including the additional twist can be written as 
\eqn\ishibashitw{\mid i\rangle\rangle= \sum_N \mid i,N\rangle \otimes U
  \Omega \widetilde{\mid i,N\rangle \period}}

The $N=2$ space-time supersymmetric
compactifications contain  two $N=2$ SCFT,
 the transverse $c=3$ and the
internal $c=9$ SCFT. Hence  $A$ and $B$
boundary conditions can be imposed 
separately on the two factors, which combination for type IIA/B
is consistent is determined by the GSO projection.

 The $c=3$ part of the SCFT given by a free complex boson and fermion 
\ceqthree\ and  will be discussed first.
As shown in \greengut\ boundary states in the light-cone gauge
impose Dirichlet boundary conditions on the light-cone coordinates
$X^+,X^-$. Hence we are dealing with D-instantons which have fixed
boundary conditions in the time direction.

When Dirichlet boundary conditions are imposed on the two transverse
coordinates $X^1,X^2$ the resulting boundary state describes a $p=-1$ 
brane in four dimensions, i.e. an event which is localized in the
four transverse directions $X^\mu=y^\mu,\mu=+,-,1,2$. Denoting
$X=X^1+iX^2$ and $\psi=\psi^1+i\psi^2$ this condition is equivalent to 
\eqn\dirichbc{\quad\big(  \partial X- \bar{\partial}X\big)\mid
  B\rangle= 0,\quad \big(\psi - i \eta  \bar{\psi}\big)\mid B\rangle =
  0 \period }

It is easy to see that with the definitions of the $N=2$ algebra given
in \ceqthree\ the D-instanton
boundary conditions correspond to the B boundary conditions \Btypebc\
for the $c=3$ system. The boundary state for a
free boson  which imposes  \dirichbc\ is constructed using  coherent
states.
Note that imposing Neuman boundary conditions for both $X^1,X^2$, also
realizes B boundary conditions. We will not discuss  D1 branes in this
paper  since new subtleties (similar to  D7-branes in ten
dimensions) arise due to the fact that there are only two transverse  
dimensions.  
One can also consider boundary conditions which impose Dirichlet
boundary  conditions on $X^1$ and Neuman boundary conditions on $X^2$.
 \eqn\dinebc{\quad \big(\partial X-\bar{\partial}X^*\big)\mid
   B\rangle=  0, \quad \big(\psi-i \eta  \bar{\psi}^*\big)\mid
 B\rangle = 0 \period }
Such boundary conditions correspond to an (Euclidean) D0-brane from
the  four dimensional perspective. As discussed in \greengut\  such a
configuration can  be related to the
standard  D0-brane by a double Wick rotation. It is easy to see that
\dinebc\ imposes $A$ boundary conditions \Atypebc\
on  the c=3 $N=2$ SCFT.

The boundary states for the internal $c=9$ SCFT are constructed by
tensoring Ishibashi states for the minimal models, subject to the
charge projection and appearance of twisted sectors. 
For A boundary conditions an Ishibashi state associated with  a primary
field of a minimal model labeled by  $l,m,s$ is given by
\eqn\aishib{\mid l,m,s \rangle\rangle = \sum_N \mid l,m,s,N \rangle
\otimes
U\Omega\mid l,m,s,N\rangle \period }
For B boundary conditions we get 
\eqn\bishib{\mid l,m,s \rangle\rangle = \sum_N \mid l,m,s,N \rangle
\otimes
U\mid l,m,s,N\rangle \period }
A  boundary state satisfying  A and B boundary conditions for the tensor
product can be constructed by the product of the boundary states \aishib\
and
\bishib\
respectively.  It is important to note that an Ishibashi state exists
 only if the primary $\Phi_{\mu,\mu}^{\lambda,\lambda}$ (for A
boundary conditions) and $\Phi_{\mu,-\mu}^{\lambda,\lambda}$ (for B
boundary conditions) appear in the partition function \partf\
 of the Gepner model.
Note that this implies that the boundary Ishibashi states satisfy the
charge projection condition and that the states in \bishib\ come from
the twisted sectors in \partf.

In \reck\  boundary states in Gepner models corresponding to A and B
boundary  conditions were  constructed by applying Cardy's  construction
 \cardy\ for each factor  of the tensor product of $n$ minimal $N=2$
theories.   
A boundary state is then labeled by a vector $\alpha=
(\lambda^\prime,\mu^\prime)$ where 
$\lambda^\prime=(l^\prime_1,\cdots,l^\prime_n)$ and $\mu^\prime=
(s^\prime_0;m_1^\prime,\cdots,s^\prime_n)$ and
given  by 
\eqn\boundstate{\mid\alpha\rangle = {1\over \kappa_\alpha}
  \sum^\beta_{\lambda,\mu}   
B^{\alpha}_{\lambda,\mu}\mid \lambda,\mu\rangle\rangle \period}
The normalization constant ${1/ \kappa_\alpha}$ 
can be determined by Cardy's condition \reck.
The factor $B^{\alpha}_{\lambda,\mu}$  is the product of
$B^\alpha_i$  in  \cardsol\
 using the modular $S$-matrix for the $N=2$ minimal models  \reck;
\eqn\bsol{B^{\alpha}_{\lambda,\mu}= e^{i\pi s_0^2/2}e^{-i\pi
    {s_0s_0^\prime\over 2}}
  \prod_{j=1}^N{\sin\big(\pi{(l_j+1)(l^\prime+1)\over k_j+2}\big)\over
   \sin^{1/2}\big(\pi{(l_j+1)\over k_j+2}\big)}
  e^{i\pi{m_jm^\prime_j\over k_j+2}}e^{-i\pi {s_js^\prime_j\over 2} \period }}

The fact that either A or B boundary conditions can be imposed for the
transverse $c=3$ and the internal $c=9$ system leads to four distinct
boundary states. Since   projection on odd integer $U(1)$ charge couples
the two sectors
in order to achieve space-time supersymmetry 
 two choices are consistent with IIA and two with IIB.
\bigskip
\begintable
 |D-1| D0 \elt
IIA|$B\otimes A$ | $A\otimes B $\elt
IIB|$ B\otimes B$ |$ A\otimes A$
\endtable
\medskip
\centerline{{\bf Table 3}: boundary conditions for type IIA/B in Gepner
models}
\bigskip
From table 3 and table 2 it is easy to see 
that the massless states appearing in the  D0 brane boundary states lie
in  vector multiplets. This fact  is
in agreement with  the interpretation of D0-branes as charged black holes
in $N=2$
supergravity. For the  D-instanton the  massless components of the
boundary state 
lie in 
hypermultiplets which means that the instanton provides a source for
the charge associated with the shift of RR-scalars, in analogy with the
D-instanton in ten dimensions \greengutb.

It is important to note  that the solution \boundstate\ is constructed by
tensoring   the Ishibashi states where 
all minimal models satisfy either A or B boundary conditions, 
hence no mixed boundary conditions are allowed. There are possible 
 generalization since the conditions \Atypebc\ and \Btypebc\  only have to
be satisfied for
the currents of the $c=9$ SCFT  which are  sums of the currents of
the minimal models. 
Hence if there is an automorphism $\cal{V}$ of the
right-moving algebra which leaves the currents invariant 
\eqn\invaut{{\cal V} \bar{T} {\cal V}^{-1}=\bar{T}, \quad {\cal V}
\bar{G}^{\pm}{\cal V}^{-1}=
  \bar{G}^{\pm},\quad {\cal V} \bar{J} {\cal V}^{-1}=\bar{J} \period}
but does not leave the each individual current
$\bar{T}_i,\bar{G}^{\pm}_i,\bar{J}_i$
of the minimal models invariant, a more general Ishibashi state can be
constructed by
\eqn\moregen{\mid \lambda,\mu\rangle\rangle= \sum_N \mid
  \lambda,\mu,N\rangle \otimes U{\cal V} \mid
  \lambda,\mu,N\rangle \period }
In the Gepner model a $B$ 
boundary state can  contain such an
Ishibashi state only if the field 
$\Phi^{\lambda,{\cal
V}\lambda}_{\mu,-{\cal V}\mu}$ appears in
the twisted sector of the partition function. One example of such an
automorphism of the $N=2$ algebra is  a
permutation of minimal models with the same $k$. 
It might also be possible to apply the more general ideas of 
\fuchs\ in this context. It would be interesting 
to analyze such boundary states further, but for the rest of this
paper the boundary states defined by \boundstate\ will be used.

\newsec{$c=3$ Gepner models and $T^2$ compactifications}
The simplest Gepner models arise for $c_{int}=3$ and correspond to
compactifications on special $T^2$. There are only three cases denoted by
$(k=1)^3$,$(k=1,k=4)$ both of which correspond to an $SU(3)$ torus and
$(k=2)^2$ which gives an $SU(2)^2$ torus.\foot{Note that unlike in the
heterotic compactification in the case of type II compactification these
tori do not lead to an enhancement of gauge symmetry.}
In this section we will treat  the $(k=2)^2$
Gepner model  in detail. The $c=3$ Gepner models are somewhat trivial, but
their simplicity makes explicit computations easier and some of the
general aspects of Gepner models are already manifest in these cases.
\subsec{torus partition function}
The boundary state $\mid B \rangle=\mid B \rangle_{osc}\times \mid B
\rangle_0 $ for a D-brane compactified on a $T^2$ contains an oscillator
part $\mid B \rangle_{osc}$ and an zero mode part $\mid B \rangle_0$ which
contains a  sum over momenta and winding modes. 

In ten dimensions the  cylinder partition
function for a supersymmetric D-brane is given by
\eqn\cylpartf{Z(q)=\langle B\mid q^{{1\over
      2}(L_0+\bar{L}_0-c/12)}\mid B \rangle \period}
The oscillator part is given by a summing over 
the $NS\otimes NS$ and $R\otimes R$ sector of the boundary state
together with the insertion of a GSO projection operator $1/2(1+(-1)^F)$; 
\eqn\cylpartfa{Z_{osc}(q)= {1\over
\eta^8}\big(\chi_v^{(8)}-\chi_c^{(8)}\big) \period }
Here the characters of the $SO(2d)$ current algebra are given by
\eqn\charsod{\chi_o^{(2d)}= {1\over 2\eta^{d}}\Big
    ( \theta_3^{d}+\theta_4^{d}\Big),\quad 
\chi_v^{(2d)}= {1\over 2\eta^{d}}\Big
( \theta_3^{d}-\theta_4^{d}\Big),\quad 
\chi_{s/c}^{(2d)}= {1\over 2\eta^{d}} \theta_2^{d} \comma }
and  $\eta=q^{1/24}\prod(1-q^n)$ is the Dedekind 
$\eta$-function. The cylinder partition function \cylpartfa\ vanishes
because of  supersymmetry.

The zero mode part of the boundary state $\mid B \rangle_0$ is defined by
sum over a sublattice of the  momentum and winding lattice
\eqn\momenta{p_L^i={G^{ij}\over \sqrt{2}}\big
    ( m_j+(B_{jk}+G_{jk})n_k\big),\quad p_R^i={G^{ij}\over \sqrt{2}}\big
    ( m_j+(B_{jk}-G_{jk})n_k\big) \comma }
where $i,j,k=1,2$ and $G_{ij},B_{ij}$ are the metric and antisymmetric
tensor background fields on the $T^2$. The boundary state is then
defined by $\mid B\rangle_0=\sum_{p_L,p_R\in \Lambda}\mid p_L,p_R\rangle$
such that 
\eqn\bndzero{\big(p_L^i+R^i_{\, j}p_R^j\big)  \mid B\rangle_0
=0 \period}
here the matrix $R^i_{\, j}$ defines the D-brane boundary conditions and
$\Lambda$ is a maximally two dimensional sublattice of the four
dimensional lattice \momenta. For boundary conditions which impose
Dirichlet boundary conditions on one and Neumann on the other direction on
the torus the matrix $R$ takes the form 
\eqn\matrixr{R^i_{\, j} =\pmatrix{
1 & 0 \cr
0 & -1 \cr
}\pmatrix{
\cos 2\theta & \sin 2\theta  \cr
-\sin 2\theta & \cos 2\theta \cr
} \period }
Note that in general only for special values of the rotation angle
$\theta$ the conditions \bndzero\ will have nontrivial solutions.

The zero mode part of the partition function is then given by
\eqn\azerom{Z_0(q)= \sum_{p_L,p_R\in \Lambda} q^{{1\over2}(p_L^2+p_R^2)} 
\period}
A Poisson resummation of \azerom\ transforms $Z_0$ to the open string
channel.

We will now specialize on  the $(k=2)^2$ Gepner model
which is equivalent to the $SU(2)^2$ torus. This is simply the unit square
torus with 
$G_{ij}=\delta_{ij}, B_{ij}=0$. The momentum lattice \momenta\ is then 
given by
\eqn\momnlat{p^i_L={1\over \sqrt{2}}(m^i+n^i),\quad p^i_R={1\over
    \sqrt{2}}(m^i-n^i),\quad i=1,2. }
The  A boundary condition corresponding a D0 brane wrapping on the
$X^1$ cycle is defined by \matrixr\ with $\theta=0$ and the conditions on
the momenta turn out to be 
\eqn\bonmom{p^1_L\mid B\rangle = p^1_R\mid B\rangle, \quad p^2_L\mid
  B\rangle = -p^2_R\mid B\rangle \period }
The lattice $\Lambda$ is defined by \momnlat\ where $n^1=m^2=0$ and
$(n^2,m^1)\in Z^2$. The zero mode part of the cylinder partition function
is then given by 
\eqn\cylmw{  Z_0^{\theta=0}(q)=
   \sum_{m_1,n_2} q^{1/4 m_1^2}q^{1/4 n_2^2} =
   \lb \theta_3^2(q)+\theta_2^2(q) \rb \period }
After a modular transformation this turns into the zero mode part of the
open string partition function
\eqn\eopstmw{ {1 \over \eta^2 } Z_0^{\theta=0}(\tilde{q})= {1\over
\eta^2 }\big(\theta_3^2(\tilde{q})+\theta_4^2(\tilde{q})\big) \comma }
where $\tilde{q} = e^{-2\pi i/\tau}$.
For later comparison with the partition function of the $(k=2)^2$ Gepner
models another boundary condition, which is given by \matrixr\
with $\theta=\pi/4$, will be important. 
Repeating the analysis above in this case it follows
that the open string partition function is given by 
\eqn\eopstmwb{{1 \over \eta^2 } Z_0^{\theta=\pi/4}(\tilde{q})= 
   {1 \over \eta^2 } \theta_3^2(\tilde{q}) \period }

The cylinder partition function is the product of the oscillator \cylpartfa\
and the zero mode part \azerom. 
In order to compare it with the open string partition
function for the Gepner model, we make 
a modular transformation and decompose the oscillator part 
in terms of $SO(6)\times SO(2)$ characters. Then we get
\eqn\modtrcyl{Z= {1 \over 2} 
  \Big({\chi_o^{(6)}\over \eta^6}{\theta_3-\theta_4\over
\eta^3}+
{\chi_v^{(6)}\over \eta^6}{\theta_3+\theta_4\over
\eta^3}-{\chi_s^{(6)}\over \eta^6}{\theta_2\over
\eta^3}-{\chi_c^{(6)}\over \eta^6}{\theta_2\over \eta^3}\Big)
Z_0^{\theta=0,\pi/4}(\tilde{q}) \period}
The partition functions for $ \theta = n \pi/2 (+ \pi/4) $ are
given by $Z_0^{\theta=0 (\pi/4)}$.
\subsec{Gepner model}
The open string partition function \modtrcyl\ should be compared with
one from the boundary states corresponding to
A boundary conditions in the Gepner model,  
\eqn\gepmdcyl{Z(q)= \langle \alpha \mid q^{{1\over
      2}(L_0+\bar{L}_0-c/12)} \mid \alpha\rangle= \sum_{\lambda,\mu}
^\beta
  B^\alpha_{\lambda,\mu} B^\alpha_{\lambda,-\mu}\chi^{\lambda}_\mu(q) \period}
A modular transformation into the open string channel gives 
\eqn\modgeptr{Z(\tilde{q})= \sum_{\lambda,\mu} ^\beta
\sum^{\rm ev}_{\bar{\lambda},\bar
     {\mu}}B^\alpha_{\lambda,\mu}
   B^\alpha_{\lambda,-\mu}S^{\lambda,\mu}_{\bar
{\lambda},\bar{\mu}}\chi^{\bar{\lambda}}_{\bar{\mu}} \comma}
where $\sum^{\rm ev}$ stands for the constraints $ l_i + m_i + s_i = 2 Z $.
This expression can be evaluated using the explicit form of
$B^\alpha_{\lambda,\mu}$ and the modular matrix $S^{\lambda,\mu}_{\bar
{\lambda},\bar{\mu}}$ for the Gepner models. The result can be found in
\reck. Here we only need the result with the same boundary conditions on
both ends of the cylinder, 
\eqn\reckres{Z_{\alpha\alpha}(\tilde{q}) =
\sum_{\bar{\lambda},\bar{\mu}}^{\rm ev} 
\sum_{v_0=0}^{K-1}\sum_{v_1,\cdots,v_n=0,1}(-1)^{\bar{s}_0}
\delta^{(4)}_{\bar{s}_0,2+v_0+2\sum v_i}
\prod_{j=1}^{n} N_{l^\prime_j}^{\bar{l}_j}
\delta^{(2k_j+4)}_{\bar{m}_j,v_0}\delta^{(4)}_{\bar{s}_j,
v_0+2v_j}\chi^{\bar{\lambda}}_{\bar{\mu}}(\tilde{q}) \comma }
up to a factor.
$ \delta^{(k)}_{m,n} $ are non-zero for $ m= n$ (mod $k$). 
$N^{l_2}_{l_1}$ is the matrix appearing in the fusion rules
among the primaries with spin $l_{1,2}/2 $ in the  $SU(2)_k$ WZW model; 
$ \phi_{l_1/2} \times \phi_{l_1/2} 
\sim \sum_{l_2} N^{l_2}_{l_1} \phi_{l_2/2} $. Namely, $ N^{l_2}_{l_1} = 1 $ 
for $ 0 \leq l_2 \leq \min (2l_1, 2k-2l_1) $ and otherwise vanishing.  
Note that the open string partition function  $Z_{\alpha\alpha}$ for two
identical D-branes only depends on 
$\lambda=(l^\prime_1,\cdots,l^\prime_n)$ in
$\alpha=(\lambda^\prime,\mu^\prime)$.

We will now specialize in the $(k=2)^2$ case. Then the matrix 
$N^{\bar{l}_j}_{l^\prime_j}$ is a $3\times 3$ matrix given by
\eqn\nmatrix{N_{{l}^\prime_j}^{\bar{l}_j}=
  \delta_{\bar{l}_j,0}+\delta_{{l}^\prime_j,1}\delta_{\bar{l}_j,2} \comma }
and hence the open string partition function depending on $l_1,l_2$ 
is given by 
\eqn\gepopenptf{Z_{l^\prime_1,l^\prime_2}(\tilde{q})=\sum_{v_0,v_1,v_2}
N_{l^\prime_1}^{\bar{l}_1}
N_{l^\prime_2}^{\bar{l}_2}(-1)^{v_0}\chi^{2+v_0+2v_1+2v_2}
\chi^{\bar{l}_1}_{v_0,v_0+2v_1}\chi^{\bar{l}_2}_{v_0,v_0+2v_2} \period}
Here $\chi^{s_0}$ denotes the $SO(6)$ character 
and $\chi^{l}_{m,s}$ is the
character of the $k=2$ minimal model corresponding to the primary
$\Phi^l_{m,s}$. Note that the form of \nmatrix\ implies that $l^\prime_i=0$ 
and $l^\prime_i=2$ give the same partition function. 
Using the field identification
for the characters $\chi^l_{m,s}=\chi^{2-l}_{m+4,s+2}$ it is also easy to
see that there are essentially only two choices of $(l^\prime_1,l^\prime_2)$ 
for non-vanishing partition functions, namely,  (a)
$(l^\prime_1,l^\prime_2)=(0,0)$ and (b) 
$(l^\prime_1,l^\prime_2)=(1,0)$.\foot{
$ (l^\prime_1,l^\prime_2 ) = (1,1) $ gives the partition function twice 
that of $ (l^\prime_1,l^\prime_2 ) = (1,0) $.  
}
   
To compare the Gepner model
partition function \gepopenptf\ with the toroidal one \modtrcyl\ it is
only
necessary (because of supersymmetry) to compare   $\chi_o^{(6)}$ part of
\modtrcyl\ with the $v_0+2v_1+2v_2=2$ (mod $4$) part of \gepopenptf.
We denote this part by $ \chi_o^{(6)} Z^o_{l^\prime_1,l_2^\prime} $.
After some calculation, we then find that  
\eqn\zktwo{\eqalign{ Z^o_{0,0} (\tilde{q}) 
   & = {1 \over 4 \eta^3} (\theta_3-\theta_4) Z_0^{\theta=0} (\tilde{q})
   \comma \cr
  Z^o_{1,0} (\tilde{q}) 
   & = {1 \over 2 \eta^3} (\theta_3-\theta_4) Z_0^{\theta=\pi/4} (\tilde{q})
   \period
}}
We relegate some details of the calculation to appendix A.
This gives the identification between 
our algebraically constructed boundary states
and D-branes wrapping around geometrical cycles;
case (a) represents the `short' branes along $ \theta = n \pi/2$
whereas case (b) represents the `long' branes along 
$ \theta = \pi/4 + n \pi/2$.

Furthermore, similar argument can be applied to other $ c= 3 $
Gepner models. (See also appendix A.) In the $ (k=1)^3 $ case,  we have
only one independent choice given by 
$ (l^\prime_1,l_2^\prime,l_3^\prime) = (0,0,0)$. It turns out that 
the corresponding boundary state represents the D-brane wrapping around 
$ \theta = n \pi /3 $ cycles of the $ SU(3)$ torus. In the $(k=1,k=4)$ case, 
we have three independent choices, 
$(l^\prime_{k=1},l^\prime_{k=4}) =$ (a) $(0,0)$, (b) $ (0,1) $ and 
(c) $ (0,2) $.
Case (a) gives the `short' D-branes wrapping around $ \theta = n \pi/3 $
while case (b) gives the `long' D-branes around $ \theta = \pi/6 + n \pi/3$.
The partition function for case (c) is the sum of the partition functions 
for case (a) and (b). The geometrical interpretation of the last case
is not completely clear. 

For the $(k=2)^2$ and $ (k=1)^3 $ case, we know explicit relations
between the space-time bosons and fermions and the free fields 
realizing the minimal models. Using them we have discussed the geometrical
interpretation of the boundary conditions for the space-time fields
in appendix B.
The above results are consistent with the possible types of D-branes
from the open string channel argument (appendix B.1).
From the closed string channel argument (appendix B.2), 
the boundary conditions for space-time fields are given
by $ m^\prime_i $ in $ \alpha $ and independent of $ l^\prime_i $.
This is complementary to the results in this section which are 
independent of $m_i^\prime$; we can choose $ l_i^\prime$ and $ m_i^\prime $
which are compatible with the two arguments. This implies 
that there may be some selection rules for allowed parameter $\alpha$.
In principle, the consistency conditions of the string theory 
such as sewing constraints (see, e.g., \reck) may give the solution.  

\newsec{Conserved and broken supersymmetry charges}
For definiteness we will consider a boundary state representing a
D-instanton in IIB compactification, which means that $B$ type boundary
conditions are imposed on the internal theory. The boundary state $\mid
\alpha \rangle$
is  labeled by $\alpha=(\lambda^\prime ,\mu^\prime)$. The fact that the
boundary state 
 corresponds to a D-brane wrapping a supersymmetric cycle
implies that four of the eight four dimensional supersymmetries are
unbroken by the boundary state.   Since the boundary state $\mid
\alpha \rangle$ contains only states with $q=-\bar{q}$  we have to
consider combinations 
$Q_L- \, e^{-i\phi} Q_R $ and $Q_L^\dagger
+ \, e^{i\phi} Q_R^\dagger$ as the unbroken supersymmetry. 
Here $e^{i\phi}$ is a phase which can be absorbed into the definition of
$Q_R$  but will be kept in the following. 
Acting with $Q_L$ shifts the left $\mu$ by $\beta_0$;
\eqn\susyone{\eqalign{Q_L \mid \alpha\rangle&= \sum
B^\alpha_{\lambda,\mu}Q_L \mid \lambda,\mu;\lambda,-\mu\rangle\rangle\cr
&= \sum B^\alpha_{\lambda,\mu}\mid \lambda,
\mu+\beta_0;\lambda,-\mu\rangle\rangle \comma }}
 up to a factor.\foot{Some states in the second line vanish.
} Here 
we have explicitly denoted the left- and right- primaries in the
Ishibashi states by $ \mid \lambda,\mu;\lambda,-\mu\rangle\rangle $.  
On the other hand acting with $Q_R$ on $\mid \alpha\rangle $ shifts the
right-moving $\mu$ by  $\beta_0$;
\eqn\susytwo{\eqalign{Q_R\mid \alpha\rangle&= \sum
B^\alpha_{\lambda,\mu} Q_R 
\mid \lambda, \mu;\lambda,-\mu\rangle\rangle\cr
&= \sum B^\alpha_{\lambda,\mu}(-1)^{s_0-1/2} \mid \lambda,
\mu;\lambda,-\mu+\beta_0\rangle\rangle \period}}
The factor $(-1)^{s_0}$ accounts for the fact that the right-moving
supercharge $Q_R$ acting on the fermionic part of the boundary state
pick up an  extra minus sign. The supercharge picks up the other factor 
$ -i $ when it goes though $ U $ in front of the right states 
(see also the appendix).
 
Shifting the summation variable $\mu$ in the second line of \susytwo\ and
using the form of $B^\alpha_{\lambda,\mu}$ we can show that
\eqn\bresult{B^\alpha_{\lambda,\mu+\beta_0}=
  B^\alpha_{\lambda,\mu} i (-1)^{s_0}e^{i\pi 2\beta_0\bullet \mu^\prime} 
\period}
Hence condition that $Q_L- \, e^{-i\phi} Q_R$ is killed by the boundary
state  $\mid \alpha \rangle$ relates the phase $\phi$ to the $U(1)$
charge  of $\alpha$; 
\eqn\omegresult{\phi= \pi  Q_\alpha= 2 \pi \beta_0 \bullet \mu^\prime \period}
For one boundary state $\mid \alpha\rangle$ such a phase $\omega$ can
be absorbed into the definition of the supercharges. Hence for any
boundary state $\mid \alpha\rangle$ constructed by the procedure in
section 3  four unbroken supersymmetries can be found. 
The importance
of  this phase becomes clear when we consider more than one boundary
state.   The boundary states $\mid \alpha_1\rangle$ and $\mid
\alpha_2\rangle$    are mutually supersymmetric only  if the two
associated
phases are related by
\eqn\susyres{Q_{\alpha_1}=Q_{\alpha_2}+2Z \period}
This is the same condition derived in \reck\   by demanding that the
open string partition function is space-time supersymmetric, i.e. satisfies
the  open string $U(1)$ projection.
Since the supercharges are space-time fermions, the relative phases
appearing in the boundary conditions may have geometrical meaning.
In the simple three cases with $c_{int} = 3$, the 
interpretation of those phases is in agreement
with the discussions in section 5 and appendix B. 
\newsec{Orbifolding and mirror map for boundary states}

The mirror automorphism \mirraut\ maps a $N=2$ SCFT into an equivalent
one by  reversing the
right-moving  $U(1)$ charges.  Although this operation is rather simple
on  the level of the conformal field theory it is the basis of mirror 
symmetry \greenepless,\candelas.

Since the string compactification is given by  a product of a $c=3$ and
$c_{int}=9$
$N=2$ SCFT for $D=4$, 
we firstly discuss the `mirror-map' for the $c=3$ system where it
is simply realized as a T-duality. For the free boson and fermion system
the mirror map can be realized in the following way
\eqn\mircthree{\bar{\partial}X \to {\bar{\partial}}X^*,\quad \bar{\psi}\to
\bar{\psi}^* \period}
Such a transformation is  T-duality in the $X^2$
direction since $\bar{\partial}X^1\to\bar{\partial}X^1$ and
$\bar{\partial}X^2\to-\bar{\partial}X^2$. This operation was named
`c-map'  in \cecotti. The  T-duality relates IIA on $R^3\times
S^1\times {\cal M}$ to IIB on $R^3\times \tilde{S}^1\times {\cal M}$
where $S^1$ and $\tilde{S}^1$ denotes circles of radius $R$ and
$\alpha^\prime/R$
respectively. Note that this map leaves the compactification manifold
$\cal{M}$ untouched. This map provides a relation between the special
K\"ahler
manifold of the  vector and quaternionic manifold of the
hypermultiplet moduli space 
since after a compactification on $S^1$ both a vector
and  hyper multiplet contain four scalars and a T-duality on $S^1$
relates  the two (the universal hypermultiplet is mapped to the
gravity 
multiplet).  
The map \mircthree\ transforms the boundary conditions  \dirichbc\ 
 into \dinebc\  and vice versa
, hence
it  maps D-instantons of type II A/B into  D0 branes of type II B/A
respectively.  Such a mapping between four dimensional
D-instantons and four dimensional black holes has been discussed on
the level of supergravity solutions in \lust.

It is well known that every $N=2$ minimal model at  level $k$ has a
 discrete $Z_{k+2}\times Z_2$ symmetry. One can use the $Z_{k+2}\times
 Z_2$
 symmetry  to define a twisted character in the following way
\eqn\twistone{Z[x,y,a,b]= \sum_{l,m,s,  \bar{m}=m-2y,\bar{s}=s-2b}
 e^{-2\pi i x ({m+\bar{m}\over  2(k+2)})} e^{2\pi i a {s+\bar{s} \over
 4}}\chi^l_{m,s}\chi^{*\; l}_{\bar{m},\bar{s}} \period}
Here the twists are labeled by $x,y=0,\cdots,k-1$ and $a,b=0,1$.
It is easy to see that the orbifolded partition function
\eqn\twisttwo{Z_{orb}= {1\over 2(k+2)}
 \sum_{x,y=0}^{k-1}\sum_{a,b=0,1}Z[x,y,a,b]=
\sum_{l,m,s}\chi^l_{m,s}\chi^{*\; l}_{-m,-s} }
is equal to an isomorphic partition function where the signs of the
 right-moving $m$ and $s$  are reversed.

The orbifolding procedure removes states from the partition function
and adds new twisted sectors. This means that in general Ishibashi
states implementing A boundary conditions get removed, whereas new
Ishibashi states for  B boundary conditions will appear. The end
result  of the orbifolding is that boundary states $\mid
\alpha\rangle$  for A and B boundary conditions get exchanged.

In the full Gepner model matter is complicated by the fact that the
$\beta$-projections involve already an orbifoldization.  In order to
preserve  the space-time supersymmetry one has to pick a subgroup of 
$\prod_i  (Z_{k_i+2}\otimes Z_2)$ which preserves the projection on odd
integer charge. This  process has been discussed in detail in
  \greenepless\  and the result is given by the mirror map which
  reverses  the right-moving $U(1)$ charges and maps \aishib\ into \bishib.

Note that the mirror map has to map a type IIA into a type IIB
configuration and it does not change the boundary condition for
the transverse $c=3$ part. Geometrically A boundary conditions
correspond to branes wrapped on middle dimensional supersymmetric
cycles, whereas B boundary conditions correspond to branes wrapped on
even dimensional supersymmetric cycles \oogurioz. The mirror map provides
a mapping of $H_3({\cal M})$ to $\sum_{i=0}^3 H_{2i}({\cal M})$. In the
Gepner
model both boundary states are characterized by the matrix
$B^\alpha_{\lambda,\mu}$ with the same $\alpha$. Hence given a geometric
interpretation of the boundary states labeled by $\alpha$ 
establishes an explicit realization of this map. 
\medskip

\ifig\fone{Relations between different D-branes in four dimensions}
{\epsfbox{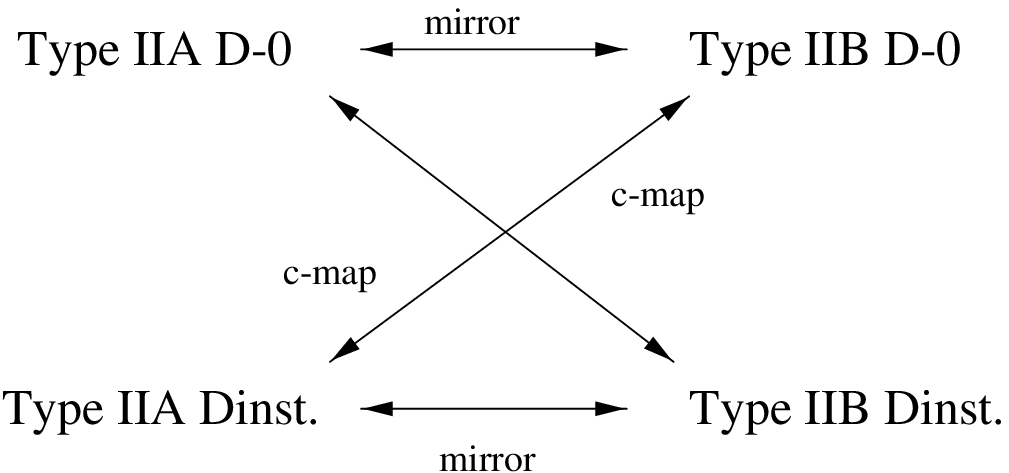}}

From the geometric point of view mirror symmetry is much more
 nontrivial than the c-map since it relates compactification manifolds of
different topology since the Hodges numbers get exchanged. On the level of
the conformal field
 theory they are the same thing applied to the two factors of the
 theory.

 \newsec{The  quintic and  $(k=3)^5$ Gepner Model}
The first example for the equivalence of a Gepner model and Calabi-Yau
compactification given in \gepner\  is the $(k=3)^5$ model which
corresponds to the quintic hypersurface in $CP^4$ defined by
\eqn\quintic{z_1^5+z_2^5+z_3^5+z_4^5+z_5^5=0 \period}
The analysis of the massless spectrum of the $(k=3)^5$ model shows
that the $(c,c)$ ring contains $101$ elements, whereas the $(a,c)$ ring
is  one-dimensional. This is in agreement with the  Betti numbers 
$h_{2,1}=101$ and
$h_{1,1}=1$ of the quintic hypersurface.  All the  information of the
boundary state $\mid
\alpha\rangle$  in  the Gepner model 
 is encoded in the matrix $B^\alpha_{\lambda,\mu}$ defined in
 \bsol. 
 
As discussed in section 7 a  minimal model of  level $k$ 
(with diagonal
partition function) has a discrete symmetry group  $Z_{k+2}$. If more
than one factor has  the
same value of $k$  there is  in addition a permutation symmetry 
 of these factors. The discrete symmetry for the $(k=3)^5$ is
therefore $(S_5\times Z_5^5)/Z_5$. In the Gepner model the quotient by
$Z_5$ is generated by $2\beta_0$.
The quintic has  the same set of discrete symmetries which is
 generated by the permutation of $z_i$ and  the transformations $z_i
 \to e^{2\pi in_i/5}z_i, i=1,\cdots, 5$.  Note that an overall phase $z_i\to
\lambda z_i,i=1,\cdots,5$ is immaterial. 

An element of discrete symmetry group $g\in G$ acts on the boundary state
in a certain way $g\mid\alpha\rangle= \mid g\alpha\rangle$, i.e., the
symmetry maps the boundary state into a new one which is characterized
by the vector $g\alpha$.
An element of the $j$-th $Z_{5}$ in $G$ acts on a primary field
 $\Phi^{\lambda,\lambda}_{\mu,\mu}$ in the following way 
\eqn\actdisc{g\Phi^{\lambda,\lambda}_{\mu,\mu}= e^{{2\pi i\over 5}
 m_j}\Phi^{\lambda,\lambda}_{\mu,\mu} \period}
Hence the action on the boundary state is given by 
\eqn\disbdstate{\eqalign {g\mid \alpha \rangle &= \sum
 B^\alpha_{\lambda,\mu} g\mid \lambda,\mu\rangle \rangle \cr
&= \sum
 B^\alpha_{\lambda,\mu} e^{{2\pi i\over 5}
 m_j} \mid \lambda,\mu\rangle \rangle \cr
&=  \sum
 B^{g\alpha}_{\lambda,\mu}  \mid \lambda,\mu\rangle \rangle \period}}
Using \bsol\ one can show that  $g\alpha$ is obtained by shifting
$m^\prime_j\to m^\prime_j+2$.

 A permutation $\pi\in
 S_5$ is acting in the following way on $(\lambda,\mu)$,
\eqn\permute{\eqalign{g_\pi(l_1,\cdots,l_n)&=
(l_{\pi(1)},\cdots,l_{\pi(n)}),\cr
 g_\pi(s_0;m_1,\cdots,m_n;s_1,\cdots,s_n
 )&=(s_0;m_{\pi(1)},\cdots,m_{\pi(n)};s_{\pi(1)},\cdots,s_{\pi(n)} )
  \period }}
Using  \bsol\ it is easy to show that $g_{\pi}\alpha$ is 
\eqn\permb{g_\pi\alpha=\big ( (l^\prime_{\pi^{-1}(1)},
\cdots,l^\prime_{\pi^{-1}(n)}),
    (s^\prime_0;m^\prime_{\pi^{-1}(1)},\cdots,m^\prime_{\pi^{-1}(n)};
s^\prime_{\pi^{-1}(1)},\cdots,
s^\prime_{\pi^{-1}(n)})\big) \period}

The discrete symmetry group $G$ acts on the set of boundary states
 labeled by $\alpha$. Thus identifying these transformation with  the
 discrete symmetries of the quintic hypersurface provides some
 information about the geometric interpretation of the boundary states
 $\mid \alpha \rangle$. Note that applying the open string partition
function in the $c_{int}=9$ case it follows that the choice 
of $\mu^\prime$ does
not influence the spectrum of open strings stretched between two identical
branes. This is in agreement with the interpretation of the $Z_{k+2}$
symmetries acting on $m^\prime_i$ as geometric transformations which change 
just the
orientation of the brane. 


  For the $T^2$ compactification, we can read off the geometrical
meaning of $m_i^\prime $ as discussed in appendix.
Since $m_i^\prime \to m_i^\prime + 2$ under 
the $Z_{k+2}$ symmetries, 
they rotate the `short'(`long') branes to other `short'(`long') 
branes. This confirms the above geometrical interpretation for 
the quintic case.    

\newsec{D-instanton effects in four dimensions}

D-instantons in the four noncompact dimensions are realized as 
wrapped  Euclidean
D-branes. From table 3  it is easy to see that for  type IIA
compactifications
D-instantons are given by  internal euclidean  D2
branes  wrapping middle dimensional cycles 
and for type IIB D-p branes wrapping  $p+1$
dimensional  cycles with $p
=-1,1,3,5$.
The boundary state formalism in the light-cone gauge is well suited for
the description of such euclidean wrapped branes \greengut.
The D-instanton is localized at a space-time point $Y^\mu$ in the four
noncompact dimensions and it is convenient to express the $Y$
dependence of the full boundary state via Fourier transformation
\eqn\boundsy{\mid B,Y\rangle= \int d^4p \ e^{ip Y} \mid B,p\rangle \period}
The inclusion of D-instantons give non-perturbative corrections to certain
amplitudes in string theory whose presence are  often demanded by
dualities. 

Wrapped euclidean branes were first discussed by Becker, Becker and
Strominger \bbs.  For Calabi-Yau compactifications of type II strings 
D-instantons correct  the metric of the
hypermultiplet moduli space. 
An important example of such an effect is
IIA near conifold  where large instanton correction due to euclidean
wrapped D2
branes smooth out the
classical singularity  in the moduli space \oogurivafa. 

The fact that a  D-instanton  breaks half  of the
eight space-time supersymmetries which are left unbroken by the
compactification implies that amplitudes in the presence of a
D-instanton are nonzero only if the fermionic zero modes associated
with the broken supersymmetries are soaked up. Very similarly to the
case of D-instantons in ten dimensional IIB string theory discussed in
\greengutb\ this leads to new instanton induced 
 t'Hooft like terms in the effective action. In the case of euclidean
wrapped
D-branes on CY four of the eight supersymmetries are broken by a
D-instanton and hence the simplest instanton term corresponds to a
four fermion term \bbs.

In the following we shall consider type IIA strings for definiteness.
Table 3 shows that this corresponds to A boundary conditions for the
internal $c=9$ SCFT.
The eight supersymmetry charges can be expressed in
the light-cone gauge notation as
\eqn\allsusy{Q^1_{\pm}={1\over \sqrt{2}}(Q\pm i\bar{Q}^\dagger),\quad \, \,
  Q^2_{\pm}={1\over \sqrt{2}}(Q^\dagger\pm i \bar{Q}),
 \quad \, \, S_{\pm}^1={1\over \sqrt{2}}(S\pm i\bar{S}^\dagger),\quad \, \,
S^2_{\pm}={1\over \sqrt{2}}(S^\dagger \pm i\bar{S}) \comma}
where the linearly and nonlinearly realized supercharges are defined in
\superch\ and \supchnl. 
The four unbroken supersymmetries which annihilate the boundary state  
are given by 
$Q_+^1,Q_+^2,S_{+}^1, S^2_{+}$ whereas the broken
supersymmetries are given by $Q_-^1,Q_-^2,S_{-}^1,
S^2_{-}$.\foot{We have absorbed phases such as 
$i \, e^{-i \phi}$ in section 6 into the definition of the right 
supercharges.}  Note that hermitian conjugation relates 
the broken and unbroken supersymmetries, i.e. $(Q_+^1)^\dagger=Q_-^2$ etc, 
which is consistent with the fact that this operation maps instantons 
into anti-instantons.
The broken supersymmetries are associated with  the
fermionic
collective coordinates in the instanton background 
 and they are related to the bosonic collective
coordinates (given by the translation of the instanton-coordinate) by the
unbroken supersymmetries 
\eqn\susyrel{\eqalign{Q_+^1 S_-^2 \mid B \rangle
&= \oint {1\over 2}\big(\partial X
  +\bar{\partial}X\big) \mid B \rangle = i {\partial \over \partial
Y^*}\mid
  B \rangle \comma \cr
Q_+^2  S_-^1 \mid B \rangle&= \oint {1\over 2}\big(\partial X^*
  +\bar{\partial}X^*\big) \mid B \rangle = i {\partial \over \partial
Y}\mid
  B \rangle \comma \cr
Q_+^2  Q_-^1\mid B \rangle&= p^+ \mid B \rangle = i {\partial
  \over \partial Y^-}\mid B \rangle \comma \cr
S_+^2  S_-^1 \mid B \rangle&= p^- \mid B \rangle = i {\partial
\over \partial Y^+}\mid
  B \rangle \comma 
}}
and similarly for other combination of broken and unbroken supercharges.
Note that the derivative $i\partial/\partial Y^\mu$ with
respect to the instanton coordinate $Y^\mu$ induces an infinitesimal
 translation and hence corresponds to the bosonic zero modes.
The broken supersymmetries are combined  with
wave functions; $\eta^1 S_-^1,\eta^2 S^2_-, \epsilon^1 Q_-^1,  
\epsilon^2 Q_-^2$.  The integration over the fermionic collective 
coordinates is then given by
$\int d^2\epsilon d^2\eta$. Instanton induced interactions are given
to lowest order in the string coupling constant 
by disk diagrams with one closed string vertex and
several broken supersymmetry generators inserted. The simplest such
term would be a four fermion term with four disk (This is the analog
of the sixteen fermion term in ten dimensional IIB superstrings
\greengutb). The inclusion of disconnected diagrams is necessary for
consistency of string perturbation theory in an D-instanton background
\polchdint,\mbggas. Such diagrams are lowest order processes and 
the experience from ten dimensional D-instantons and special examples 
where exact instanton contributions are available in four dimensions 
\oogurivafa\ indicate that there are contributions from higher charged 
D-instantons and an infinite number of higher order perturbative 
fluctuations around the D-instanton. Such perturbative fluctuations around 
the D-instanton appear as higher genus worldsheets with D-instanton boundary 
conditions in this framework. 

An $N=2$ hypermultiplet contains on shell four fermionic and four bosonic
states. In the light-cone gauge one can decompose a Dirac spinor $\Psi$
into two two-component spinors $\psi$ and $\dot{\psi}$ which satisfy
$\Gamma^+ \psi=0$ and $\Gamma^-\dot{\psi}=0$.

The fermionic states in the hypermultiplets can be created by acting with
the 
supersymmetry charges on the scalar $\mid \lambda, \mu;
\lambda,\mu\rangle$. Here $\lambda, \mu$ denote a scalar state in the
NS-NS sector with $s_0=\bar{s}_0=0$ and hence $q=\bar{q}=\pm 1$. 
 It is understood that this expression includes the momentum
dependence.  There are
$2h_{2,1}+2$ such states in a Gepner model corresponding to a Calabi-Yau
compactification with Betti number $h_{2,1}$ and we shall use the label
$I$ to distinguish them; 
\eqn\fermwfa{ \mid \dot{\psi}^{1\;I}\rangle  =  {1\over k^+}\dot{\psi}^{1\;I}
 Q_-^1\mid \lambda, \mu;
\lambda,\mu\rangle, \quad\mid \dot{\psi}^{2\;I}\rangle = {1\over k^+}
\dot{\psi}^{2\;I}Q_-^2\mid \lambda, \mu;
\lambda,\mu\rangle \period }
The states \fermwfa\ are two physical polarizations. In order to obtain
the four fermionic states in a hypermultiplet \fermwfa, 
$(\lambda,\mu)$ has to be combined with a charge conjugate field $(
\lambda,-\mu)$. Since the label $I=1,\cdots,2h_{2,1}+2$ this gives
$h_{2,1}+1$ hypermultiplets.

The disk diagram with the insertion of a fermionic state 
 $\psi^{1,2\;I}$ and one
fermionic zero mode will be denoted $A_\psi^{1,2\;I}$ and 
can be calculated by acting with a broken
supersymmetry on the boundary state and evaluating  the overlap with
the states \fermwfa. Using \susyrel\ to evaluate the
resulting one point function for the linearly realized supercharges gives
\eqn\fourferma{\eqalign{\langle \dot{\psi}^{1\;I}\mid\epsilon^1Q_-^1 
\mid B\rangle &= 
\epsilon^1\dot{\psi}^{1\;I}\langle \lambda,\mu\mid B\rangle \comma \cr
 \langle \dot{\psi}^{2\;I}\mid\epsilon^2 Q_-^2 
\mid B\rangle &= 
\epsilon^2\dot{\psi}^{2\;I}\langle \lambda,\mu\mid B\rangle \period }}
The nonlinearly realized supercharges give
\eqn\fourfermb{\eqalign{
\langle \dot{\psi}^{1\;I}\mid \eta^1 S_-^1 \mid B\rangle &= 
\eta^1 {k\over k^+}\dot{\psi}^{1\;I} \langle\lambda,\mu\mid B\rangle = \eta^1
{\psi}^{1\;I}\langle\lambda,\mu\mid B\rangle \comma \cr
\langle \dot{\psi}^{2\;I}\mid \eta^2 S_-^{2} \mid B\rangle &= 
\eta^2 {k^*\over k^+}\dot{\psi}^{2\;I} \langle\lambda,\mu\mid B\rangle = \eta^2
{\psi}^{2\;I}\langle\lambda,\mu\mid B\rangle \period }}
 Where the Dirac equation $\Gamma^\mu p_\mu\Psi=0$  is used to relate the
$\psi$ and $\dot{\psi}$ components of the spinor;  
\eqn\diraceq{{k\over k^+}\dot{\psi^1}={\psi^1},\quad {k^*\over
k^+}\dot{\psi}^2={\psi^2} \period}
The four fermionic zero modes of the D-instanton can then be saturated by
four disk amplitudes given in \fourferma\ and \fourfermb. The result can
be expressed in a covariant form 
\eqn\fourfermint{\int d^2\epsilon d^2\eta
  A_{\psi^I}A_{\psi^{J}}A_{\psi^K}A_{\psi^{L}}=
      (\bar{\psi}^I \psi^{J})\;(\bar{\psi}^K\psi^{L})\;
      R_{I{J}K{L}} \period }
Using the fact that the overlap 
$ \langle \lambda,\mu \mid B \rangle=
B^\alpha_{\lambda,\mu}$ the tensor $R_{I{J}K{L}}$ is given in terms of the
matrix $B^\alpha_{\lambda,\mu}$ by
\eqn\rresult{R_{I{J}K{L}} = B^\alpha_{\lambda^I, \mu^I}
B^\alpha_{\lambda^{{J}}, \mu^{{J}}}B^\alpha_{\lambda^K,\mu^K}
B^\alpha_{\lambda^{{L}},\mu^{L}} \period}
Here $\lambda^I, \mu^I$ labels  the states in the hypermultiplet coming
from massless states in the NS-NS
sector of the Gepner model and we have omitted an inessential 
normalization factor. 
The  four fermion term \fourfermint\ agrees with the result in 
\bbs\ which was obtained with different methods.

Supersymmetry relates this  four fermion term to other instanton induced
terms in the effective action which can also be calculated directly using
the boundary states and the broken supersymmetry charges.
A correction of the hypermultiplet metric is produced  by two disks 
with a vertex operator for a
scalar in the hypermultiplet and two broken susy generators on each disk.

For a
massless scalar in the hypermultiplet we insert a vertex operator 
$V_{\Phi^{i}} = \Phi^{\lambda,\lambda}_{\mu,\mu}e^{ikX}$ at the center of
the disk. The scalars in the hypermultiplet
$\Phi^{i}$ are  labeled by  $i=1,\cdots,4(h_{2,1}+1)$ denoting the
scalars of the NS-NS sector with $q=\bar{q}=\pm 1$ and the scalars in
the RR
sector related by spectral flow. A disk diagram with such a state inserted
together with two broken
supercharges can be written in the cylinder frame as 
\eqn\diska{\eqalign{A_{\Phi^i}= \langle \lambda,\mu;\lambda,\mu \mid 
Q_-^1 S^2_- \mid  B\rangle &= {1\over 2} \langle
\lambda,\mu;\lambda,\mu\mid  (Q-i\bar{Q}^\dagger)(S^\dagger-i\bar{S}) \mid B\rangle\cr 
&=\langle
\lambda,\mu;\lambda,\mu\mid (QS^\dagger-S^\dagger Q)\mid  B\rangle \cr
&=  k^*\langle
\lambda,\mu;\lambda,\mu\mid B\rangle\cr
&= k^* B^\alpha_{\lambda,\mu} \period}}
Where we used the boundary conditions 
on the supersymmetry charges imposed by $\mid B\rangle$ to turn all the 
supersymmetry charges into leftmoving ones.
Here and in the following we dropped the wavefunction such as $\Phi_i$ 
associated with the scalars in the hypermultiplet for notational ease.
  On the second disk another hypermultiplet vertex operator $V_{\Phi^{j}} =
\Phi^{\bar{\lambda},\bar{\lambda}}_{\bar{\mu},\bar{\mu}}e^{ikX}$ together
with the 
 two remaining 
broken supercharges have to be inserted;  
\eqn\diskb{\eqalign{ A_{\Phi^{{j}}}=\langle
\bar{\lambda},\bar{\mu};\bar{\lambda},\bar{\mu} \mid 
Q_-^2 S_-^1\mid  B\rangle &=  \langle
\bar{\lambda},\bar{\mu};\bar{\lambda},\bar{\mu}\mid ( Q^\dagger S-S Q^\dagger) \mid B\rangle \cr
 &= k\langle
\bar{\lambda},\bar{\mu};\bar{\lambda},\bar{\mu} \mid B\rangle\cr
&=k B^\alpha_{\bar{\lambda},\bar{\mu}} \period }}
\ifig\fone{D-instanton corrections to the hypermultiplet metric
$G_{i\bar{j}}$}
{\epsfbox{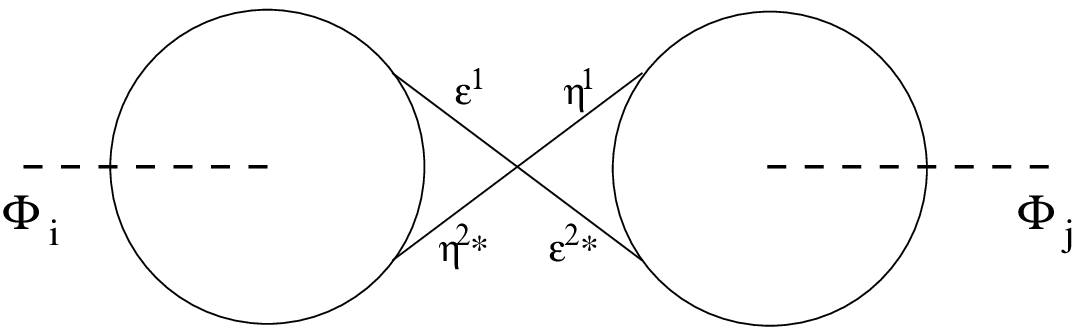}}
Putting together the two disk amplitudes \diska\ and \diskb\ and
integrating over the fermionic collective coordinates gives
\eqn\diskone{A=\int d^2\epsilon d^2\eta
  A_{\Phi^i}A_{\Phi^{\bar{j}}}= {\partial\over \partial X} \Phi^i
  {\partial\over \partial X^*}\Phi^{{j}}
B^\alpha_{\lambda,\mu}B^\alpha_{\bar{\lambda},\bar{\mu}} \period }
 Different combinations of the fermionic zero modes on the two disks
produce similar terms, i.e., when $Q_-^1$ and $Q^2_-$ are inserted  on
one  disk and $S_-^1$ and
$S^2_-$ on the other we get 
\eqn\plamp{\eqalign{A_{\Phi^i}&= \langle \lambda,\mu;\lambda,\mu \mid 
Q_-^1 Q^2_- \mid  B\rangle \cr 
&= \langle
\lambda,\mu;\lambda,\mu\mid  (QQ^\dagger-Q^\dagger Q)\mid B\rangle \cr
&= k^+ B^\alpha_{\lambda,\mu} \period }}
and  
\eqn\minamp{\eqalign{A_{\Phi^j}&= \langle
\bar{\lambda},\bar{\mu};\bar{\lambda},\bar{\mu} \mid 
S_-^1 S^2_- \mid  B\rangle \cr
&=   \langle
\bar{\lambda},\bar{\mu};\bar{\lambda},\bar{\mu}\mid  (SS^\dagger-S^\dagger S)
\mid B\rangle\cr
&= {kk^*\over k^+}\langle
\bar{\lambda},\bar{\mu};\bar{\lambda},\bar{\mu} \mid B\rangle\cr
&= k^- B^\alpha_{\bar{\lambda},\bar{\mu}} \period }}
 Integration over the fermionic zero modes like in \diskone\ then gives

\eqn\disktwo{A=\int d^2\epsilon d^2\eta
  A_{\Phi^i}A_{\Phi^{{j}}}= {\partial\over \partial X^+} \Phi^i
  {\partial\over \partial X^-}\Phi^{{j}}
B^\alpha_{\lambda,\mu}B^\alpha_{\bar{\lambda},\bar{\mu}} \period}
Putting together all these terms the instanton induced correction of the
metric can then be expressed in a Lorentz covariant form 
\eqn\instmetr{A= \partial_\mu \Phi^i \partial^\mu \Phi^{{j}}
g_{i{j}}
  e^{-S_{inst}} \period }
It is easy to see that the correction to the metric due to the lowest
order instanton process gives 
\eqn\metres{\eqalign{g_{i{j}}&= 
    B^\alpha_{\lambda,\mu}B^\alpha_{\bar{\lambda},\bar{\mu}}\cr
&= e^{i\pi {s_0^2+\bar{s}_0^2\over 2}}e^{-i\pi
    {s^\prime_0(s_0+\bar{s}_0) \over 2}}
  \prod_{j=1}^n\Big({\sin\big(\pi{(l^\prime_j+1)(l_j+1)\over
k_j+2}\big)\over
    \sin^{1/2}\big(\pi{(l_j+1)\over k_j+2}\big)}\cr
&\times\;{\sin\big(\pi{(l^\prime_j+1)
  (\bar{l}_j+1)\over k_j+2}\big)\over
    \sin^{1/2}\big(\pi{(\bar{l_j+1)\over k_j+2}\big)}}
  e^{i\pi{m^\prime_j(m_j+\bar{m}_j)\over k_j+2}}e^{-i\pi
{s^\prime_j(s_j+\bar{s}_j) 
\over 2}}\Big) \period }}
Such an instanton induced correction has to be weighted by a factor
$e^{-S_{inst}}$ where $S_{inst}$ is the action of the instanton. In the
  case of  the wrapped branes the action is simply given by $S_{inst}=
  e^{-\phi}Vol(C)$ where $C$ is the cycle on which the (euclidean)
  D-brane  is wrapped. The action can be read off from the one point
  function  of the dilaton on the disk. The dilaton corresponds to 
the scalar state in the RR sector which has $q=\bar{q}=0$ and is given by a
linear combination of the states with $s_0=\pm 1,\bar{s}_0 =\mp 1$ and
$l_i=m_i=s_i=0, i=1,\cdots,n$. The overlap
  with the boundary state is then 
\eqn\overlap{\langle s_0,0,0\mid \alpha\rangle= B^\alpha_{0,0} = {i \over
    \kappa_\alpha} \prod_j^n 
  \sin\big(\pi{(l^\prime_j+1)\over k_j+2}\big) \period}
  The instanton induced terms in the effective action are of importance
since sometimes the exact corrections can be obtained by other means like
duality \anton\ or symmetry constraints \oogurivafa. The instanton
calculations given above can then be used to determine D-instanton
partition functions \greengutd\ which in turn can be related to
complicated matrix integrals.
\newsec{Conclusions}

In this paper we have analyzed the boundary states constructed by
Recknagel and Schomerus \reck. Using the light-cone gauge description of
boundary states, there are two $N=2$ SCFT, one for the transverse  and one
for the internal degrees freedom. The consistent boundary conditions give
D-instantons and D-0 branes (black holes) corresponding to euclidean
wrapped D-branes for type IIA and IIB.
The boundary states are characterized  by a set of discrete labels
$\alpha=({\mu^\prime,\lambda^\prime})$.
The open string partition function of strings ending on identical branes
depend only on $\lambda^\prime$, whereas the relative phase of the left- and
right-moving part of the conserved supersymmetry charges depends 
only on $\mu^\prime$. This suggests existence of the selection rule
for the allowed parameter $ \alpha$
since both arguments give geometrical interpretations
and they should be compatible.
The understanding of this rule may give useful insight into 
the classification of the supersymmetric cycles.
The discrete
symmetries of the Gepner model map boundary states labeled by different
$\mu^\prime$ into each other. The construction of the D-brane boundary states
used Ishibashi states satisfying A or B boundary conditions for each
minimal model in the tensor product. It would be very interesting to
generalize this construction and to find new (`non-rational')
supersymmetric boundary states in Gepner models. 

The Gepner model (and the associated boundary states) provide one point in
the moduli space of compactifications (and branes wrapping cycles). It is
important to note that mirror symmetry \greenepless,\candelas\ was established 
in this
context and then extrapolated through the moduli space of
compactifications. The dependence of D-brane boundary states on the
K\"ahler and complex structure deformations was analyzed in \oogurioz\
using the topologically  twisted sigma model. It might be useful to
consider the topologically twisted minimal models \ttwm\  to analyze 
the behavior of Gepner
model boundary states under marginal deformations.    

The boundary states and the space-time supersymmetry charges were used to
calculate instanton induced corrections to the hypermultiplets in the four
dimensional $N=2$ effective action. This calculation uses the overlap of a
boundary state and a closed string state, corresponding to the insertion
of a closed string vertex operator on the disk.

We observed that the mirror automorphism (`c-map') of the transverse
conformal field theory is given by a time-like T-duality which maps
D-instantons into D0 branes corresponding to RR-charged black holes in 
$N=2$
supergravity. In this context the one point functions for a boundary state
can be related to the properties of the corresponding black hole. It would
be very interesting to use the exact solution provided by the Gepner model
boundary states to calculate higher point functions on the disk which
correspond to scattering off the black hole, absorption and Hawking
radiation. We hope to report progress in this direction elsewhere.

\bigskip
\bigskip
\noindent{\bf Acknowledgments}
\medskip
{We would like to thank A. Recknagel and  V. Schomerus for very useful
correspondence. We are also grateful to N. Ishibashi, A. Kato and H. Ooguri
for useful correspondence and/or conversation.
M.G. gratefully acknowledges the hospitality of the Theory Division at
CERN while this work was completed. 
The work of. M.G. was supported in part by DOE grant DE-FG02-91ER40671 
and NSF grant PHY-9157482, whereas 
the work of Y.S. was supported in part 
by Japan Society for the Promotion of Science.
}

\appendix{A}{Calculation of the partition functions}
In appendix A, we calculate 
the partition functions from the boundary states 
for the $ c=3$ Gepner models.
For this purpose, we need the characters of the level 
$k$ $N=2$ minimal model in the NS sector ($s=0,2$) \gepner,
\eqn\chrctr{\eqalign{
    \chi_{m,s}^{l} (q) &= \sum_{j=0}^{k-1} \ c^l_{m+4j-s} (q) 
           \theta_{2m+(4j-s)(k+2),2k(k+2)} (q) \comma \cr
     \theta_{M,K} (q) &= \sum_{n \in Z} q^{K \lb n+{M \over2K} \rb^2 \comma }
}}
where $c_m^l $ are Hecke modular forms depending on $k$.
We will use the identities $\theta_{M+2K,K} = \theta_{M,K}$,
$c^l_m = c^l_{-m} =  c^{k-l}_{k\pm m}  =  c^l_{m + 2k}$ and 
$ c^l_m = 0 $ unless $ l+m \in 2 Z $.
\subsec{$(k=2)^2$ case}
For $ k= 2 $, $ c_m^l$ reduce to the fermion characters
\eqn\fermion{\eqalign{
   \chi_0 & = {1 \over 2} \lb \sqrt{\theta_3/\eta} + \sqrt{\theta_4/\eta} \rb
    \comma \quad 
   \chi_{1/2}  =  {1 \over 2} \lb \sqrt{\theta_3/\eta} 
                              -\sqrt{\theta_4/\eta} \rb
    \comma \cr
  \chi_{1/16} & =  \sqrt{\theta_2/2\eta} \period
}} 
Precise relation is given by
\eqn\chc{
  \eta(q) c^0_0(q) = \chi_0(q) \comma \qquad
  \eta(q) c^2_0(q) = \chi_{1/2}(q) \comma \qquad
  2 \eta(q) c^1_1(q) = \chi_{1/16}(q) \period   
}
Also, the following combinations of $ \theta_{M,K} $ are used;
\eqn\si{\eqalign{
  S_1(q) & \equiv \theta_{0,16}(q) + \theta_{16,16}(q) =
   {1 \over 2} \lb  \theta_3 (q^2) + \theta_4(q^2) \rb \comma \cr
  S_2(q) & \equiv \theta_{-8,16}(q) + \theta_{8,16}(q) 
            = {1 \over 2} \lb  \theta_3 (q^2)  - \theta_4(q^2) \rb \comma \cr
  S_3(q) & \equiv \theta_{\pm 4,16}(q) + \theta_{\mp 12,16}(q)  
                  = {1 \over 2} \theta_2 (q^2) 
  \period   
}}

In this case, we have three independent non-vanishing 
partition functions labeled by $(l_1^\prime, l_2^\prime) = (0,0),(1,0),
(1,1) $. Using \gepopenptf\ and the duplication formulas for 
$\theta_i$,  we find that 
\eqn\zllzero{\eqalign{
  Z^o_{0,0}(\tilde{q}) & = 4 c^0_0 c^2_0 S_1 S_2 
                 + 2 \lb (c_0^0)^2 + (c_0^2)^2 \rb S_3^2 \cr
    &= {1 \over 4 \eta^3(\tilde{q})} \lb 
    \theta_3 (\tilde{q}) -\theta_4(\tilde{q}) \rb
      \lb \theta_3^2 (\tilde{q}) + \theta_4^2(\tilde{q}) \rb \period
}} 
In the $(1,0) $ case, 
we have additional contribution
\eqn\delz{
   \Delta Z  = 4 c_0^0 c_0^2 S_3^2 + 2 \lb (c_0^0)^2 + (c_0^2)^2 \rb S_1 S_2
 \period
}
Therefore the total partition function is 
\eqn\zerone{\eqalign{
   Z^o_{1,0}(\tilde{q}) 
& = Z^o_{0,0}(\tilde{q}) + \Delta Z \cr
     &=  {1 \over 2 \eta^3(\tilde{q})} 
     \lb \theta_3 (\tilde{q}) -\theta_4(\tilde{q}) \rb
      \theta_3^2 (\tilde{q}) \period 
}}
We easily confirm that the partition function of the last case 
$(1,1)$
is $ 2 Z^o_{1, 0}$.
\subsec{other cases}
The other $c=3$ Gepner models correspond to the $SU(3)$ torus.
The metric and the anti-symmetric tensor in \momenta\ are
given by $G_{11}=G_{22} = 1$ and $G_{12}=B_{12}=1/2$ in the 
coordinate system whose basis is along adjacent two roots.
By a similar argument in section 5.1, the zero-mode part of
the torus partition functions is given by
\eqn\suthree{ Z_{SU(3),0}^\theta(\tilde{q}) = \sum_{m,n \in Z} 
    \tilde{q}^{ a (m^2+n^2+mn)} \comma
}
where $ a = 1 $ for the D-branes wrapping on the cycles 
$ \theta = n \pi/3$ of the $SU(3)$ torus and $ a=1/3$ for 
$ \theta = (2n+1)\pi/6$.

In the $(k=1)^3$ case, we have only one independent partition function
labeled by $(l_1^\prime,l_2^\prime,l_2^\prime) = (0,0,0)$ 
on the Gepner model side. In addition,  $c_m^l $ reduce to 
$ \eta^{-1}$.
Then using the identity
\eqn\ident{\eqalign{& \sum_{l,m,n \in Z} q^{{1 \over 6}(2l-1)^2}
      \lbb  3 \, q^{ {1 \over 6} \lmb (6n+2l+2)^2 + (6m+2l+2)^2 \rmb}
   + q^{{1 \over 6} \lmb (6n+2l-1)^2 + (6m+2l-1)^2 \rmb} \rbb \cr 
  & \qquad = \sum_{l,m,n \in Z}  q^{2(l-1/2)^2 +m^2+n^2 +mn} \comma
}}
we obtain
\eqn\kone{
   Z^o_{0,0,0}(\tilde{q}) = 
    {1 \over  \eta^3 (\tilde{q})} 
   \lb \theta_3(\tilde{q}) - \theta_4(\tilde{q}) \rb 
   Z_{SU(3),0}^{\theta = 0} (\tilde{q}) \period 
}

In the $ (k=1,k=4) $ case, on the Gepner model side 
we have three independent partition functions labeled by 
$(l_{k=1}^\prime,l_{k=4}^\prime) = (0,0), (0,1),(0,2)$.
Since $ c_m^l$ for $ k=4 $ sector are complicated, 
it seems difficult to analytically deal with 
the partition functions. However, using computers we find that
$ Z^o_{0,0} = Z^o_{0,0,0}/2$,  
\eqn\konefour{
      Z^o_{0,1}(\tilde{q}) = {1 \over 2 \eta^3 (\tilde{q})} 
   \lb \theta_3(\tilde{q}) - \theta_4(\tilde{q}) \rb 
   Z_{SU(3),0}^{\theta = \pi/6} (\tilde{q}) \comma   
}
and $Z^o_{0,2} = Z^o_{0,0} + Z^o_{0,1}$.
\appendix{B}{Geometrical interpretation of the boundary conditions}
The $(k=2)^2$ and $(k=1)^3$ Gepner model correspond to $SU(2)^2$ and 
$SU(3)$ torus respectively. 
In these cases, one can explicitly construct the sigma-model 
variables for $ T^2 $ using free field realization. 
This allows us to find the geometrical meaning of the boundary conditions.  

The tensor product of two $k=2$ $N=2$ minimal models are realized
by two real bosons and real fermions as 
\eqn\ntwoct{\eqalign{
    T(z) &= 
      - {1 \over 2} \sum_{j=1}^2 \lb \del \varphi_j \del \varphi_j 
                                               + \psi_j \del \psi_j \rb
         \comma \cr
   J(z) &= {i \over \sqrt{2} } \del \lb \varphi_1 + \varphi_2 \rb 
             \comma \qquad 
   G^\pm (z) \ = \ {1 \over \sqrt{2} } \sum_j  \psi_j 
                    \ e^{\pm i \sqrt{2} \varphi_j} 
   \period } }
Using $\varphi_i$ and $\psi_i$, the chiral complex boson and fermion 
on the torus are given by \GS\foot{
Precisely, to get Gepner models we need to twist 
the tensor product of the minimal models (see section 3). 
The sigma-model variables discussed here and in the following are 
in the untwisted sectors.}
\eqn\pxktwo{\eqalign{
    \Psi (z) &=
        {1 \over \sqrt{2}} \ e^{i H /\sqrt{2}} 
      \comma  \cr
    \del X(z) &= 
       \sqrt{2}  
            \lb \psi_1 \ e^{-i(\varphi_1 - \varphi_2)/\sqrt{2})} 
                 + \psi_2 \ e^{i(\varphi_1 - \varphi_2)/\sqrt{2})} \rb 
   \comma } }
where $ H = \sum_j \varphi_j $.

In the $(k=1)^3$ case, the tensor product of the minimal models 
is realized by three real bosons as 
\eqn\konect{\eqalign{
     T(z) &= - {1 \over 2} \sum_{j=1}^3 
      \del \varphi_j \del \varphi_j \comma \cr
   J(z) &= {i \over \sqrt{3}} \sum_j \del \varphi_j
             \comma \qquad 
   G^\pm (z) \ = \  \sqrt{{2 \over 3}} \sum_j \ e^{\pm i \sqrt{3} \varphi_j} 
                     \period 
}}
The complex boson and fermion on the torus are (see, e.g. \Warner)
\eqn\pxkone{
   \Psi  =  \  \ e^{i H/\sqrt{3}} \comma \qquad
   \del X \ = \ {1 \over \sqrt{3}} 
           \sum_j \ e^{i(H -3 \varphi_j)/\sqrt{3}} 
   \period
}

\subsec{boundary conditions in the open string channel}

We first discuss the geometrical meaning of the boundary 
conditions from the open string channel. A and B boundary
conditions are given by
\eqn\opnbc{\eqalign{
   ({\rm A}) \qquad 
    & J_n + \bar{J}_{-n} = 0 \comma \qquad 
   G^\pm_r - \eta \ \bar{G}^\mp_{-r} = 0
    \comma \qquad  L_n = \bar{L}_{-n} \comma  \cr
   ({\rm B}) \qquad 
   & J_n -  \bar{J}_{-n} = 0 \comma \qquad 
     G^\pm_r - \eta \ \bar{G}^\pm_{-r} = 0
   \comma \qquad L_n = \bar{L}_{-n} \comma
}}
with $\eta = \pm 1$.

In the $(k=2)^2$ case, these boundary conditions are translated into
\eqn\opbcktwo{
  \varphi_j = \epsilon \bar{\varphi}_j + \pi  n_j/\sqrt{2} \comma 
  \qquad  \psi_j = (-1)^{m_j} \bar{\psi}_j
   \comma 
}
where $ \epsilon = -1$ for A and $ +1$ for B;
 $ m_j + n_j = 2 Z $ for $\eta = +1$ and $ m_j + n_j = 2 Z +1 $ 
for $\eta=-1$. In terms of the sigma-model variables, these imply
\eqn\xpbcktwo{
  \Psi =  \ e^{\pi i (n_1+n_2)/2} \bar{\Psi}^* \comma \qquad   
  \del X =  (-1)^{n_1+m_1} 
            \ e^{\pi i (n_1+n_2)/2} \delbar \bar{X}^* \comma
}
for A and similar expression with $ \bar{\Psi}, \bar{X}$ 
instead of $ \bar{\Psi}^*, \bar{X}^*$
for B.
Thus A boundary conditions represent the D-branes wrapping 
around the cycles $ \theta (=\arg X) = \pi (n_1 + n_2)/4 $. 
These cycles are the shortest and the second shortest cycles of 
the $SU(2)^2$ torus. On the other hand,  B boundary 
conditions represent the N-N or the D-D boundary conditions
in both directions of the torus.

The $(k=1)^3$ case has been discussed in \oogurioz.
Similarly to the above, we find that A and B
boundary conditions give $ \varphi_j = \epsilon \bar{\varphi}_j + c_j $,
where $ c_j = 2\pi n_j/\sqrt{3} $, $n_j \in Z $ for $\eta = +1$ and 
$2\pi (n_j+1/2)/\sqrt{3} $ for $\eta = -1$. Geometrically these imply 
\eqn\xpbckone{
   \Psi = \eta \, e^{2\pi i n/3} \bar{\Psi}^* \comma \qquad 
   \del X =  \  e^{2\pi i n/3} \delbar \bar{X}^*
   \comma 
}
for A with $ n = \sum n_j $ and 
similar expressions with $ \bar{\Psi},\bar{X}$  
instead of $ \bar{\Psi}^*,\bar{X}^* $ for B. 
Thus A boundary conditions represent the D-branes wrapping around
the cycles $ \theta = \pi  n/3$. These are the shortest cycles 
of the $ SU(3) $ torus.  B boundary conditions correspond to the 
N-N or D-D boundary conditions.
\subsec{boundary conditions in the closed string channel}

Next, we discuss the boundary conditions in the closed string 
channel, i.e., on the boundary states. They are given by
\Atypebc\ and \Btypebc.

To analyze them, we first recall that generically the $N=2$ minimal
models are realized by parafermions and a free boson \Qiu;
\eqn\prfrmn{\eqalign{
   T (z) &= T_{PF} (z) - {1 \over 2} \del \varphi \del \varphi \comma 
   \qquad J(z)  \ = \ i \gamma^{-1} \del \varphi \comma \cr
   G^+ (z) &= \sqrt{2} \gamma^{-1} \psi^{PF}_1 \ e^{i \gamma \varphi} \comma 
   \qquad G^- (z) \ = \ \sqrt{2} \gamma^{-1} (\psi^{PF}_1)^\dagger 
       \ e^{- i \gamma \varphi} \comma 
}}
where $T_{PF}$ is the energy-momentum tensor for the parafermions;
$ \gamma = \sqrt{(k+2)/k} = \sqrt{3/c} $ ; 
$k$ is the level of the minimal model.
In addition, the highest weight states corresponding to 
$ \ket{ l,m,s } $ are written as
\eqn\ntwohws{
 \Phi^l_{m,s} =  \phi^l_{m-s} \ e^{i \gamma^k_{m,s} \varphi} \comma \qquad
  \gamma^k_{m,s} = {m-s(k+2)/2 \over \sqrt{k(k+2)}} \period 
}
$ \phi^l_q $ are the primary fields in the parafermion theory and 
the above expression of $\gamma_{m,s}^k $ is valid in the standard range.
From these realizations, we find that the states in the module of 
$ \ket{ l,m,s } $, i.e. $ \ket{ l,m,s;N } $, take the form
\eqn\PFrep{
  (\hbox{PF modes}) \times (\hbox{non-zero modes of } 
  \varphi  )  
     \ket{ \gamma^k_{m,s} + n \gamma } \comma
   \qquad n \ \in \ Z \comma 
}
where $ \ket{ p } $ is the momentum eigenstate of $ \varphi $. 

For $k=1$, the `parafermions' are trivial, namely, $ \phi^l_{m-s} = 1$,
 whereas for $k=2$ they are usual fermions. 
Notice that the sigma-model fermions both in \pxktwo\ and \pxkone\
take the form 
\eqn\PsiJ{
     \Psi(z) = \ e^{i \gamma^{-1} H (z)} \comma \qquad 
    i \gamma^{-1} \del H(z) \ = \ J (z)
    \period 
}

For definiteness, we focus on the $(k=2)^2$ case and A boundary conditions
for the time being. By definition,  A boundary states satisfy
\eqn\AbcJ{
   J_n \ket{ \alpha } _A = \bar{J}_{-n} \ket{ \alpha } _A 
   \period
}
Using this and \PsiJ, we find that 
\eqn\AbceH{
   e^{i H/\sqrt{2}} (\sigma) \ket{ \alpha } _A 
    =  \ e^{- i \bar{H}/\sqrt{2}} (\sigma) \ e^{i(x+\bar{x})/\sqrt{2}} 
   \ket{ \alpha } _A
   \comma  
}
where $ x = x_1 + x_2 $; $ x_j $ are the zero modes of 
$ \varphi_j = x _j- i \alpha^j_0 \ln z + \cdots $ and 
we have set $ z = e^{i\sigma} $, $ \zbar = e^{-i\sigma}$. 
From \PFrep, we see that the zero-mode operator 
$ e^{i(x+\bar{x})/\sqrt{2}} $ acts only on the 
momentum eigenstates of $ H $ and $ \bar{H} $ at the base. Since 
\eqn\gmmktwo{
   \gamma^{k=2}_{m,s} + {1 \over \sqrt{2}} = \gamma^{k=2}_{m-2,s-2}
    \comma
}
it follows that 
\eqn\AbceHtwo{\eqalign{
   e^{i H/\sqrt{2}} (\sigma) \ket{ \alpha } _A  &=  
           \ e^{- i \bar{H}/\sqrt{2}} (\sigma) \ 
       {1 \over \kappa^A_{\alpha}} \sum_{\lambda, \mu} B_{\lambda, \mu}^\alpha
                       \kket{ \lambda, \mu - 2 \eta_0 } _A \comma \cr
    &= \ -i \eta \, e^{- i \bar{H}/\sqrt{2}} (\sigma) 
              \ e^{+2\pi i \hat{Q}_\alpha} 
                       \ket{ \alpha } _A \period 
}}
Here $ \eta_0 = (0;2,2;2,2) $, 
\eqn\hatQalph{
   \hat{Q}_\alpha \ \equiv \ 2 \eta_0 \bullet \mu' \ = \ 
       (m'_1+m'_2)/4 - (s'_1+s'_2)/2 \comma
}
and $ \alpha = (\lambda',\mu')$. 
We have also used 
$ U \, e^{i p x_j} \, U^{-1} = (-1)^{p^2/2} e^{-i p x_j}$,
$ U \, c \, U^{-1} = c^*$ for a $c$-number, 
$ -i \eta = (-1)^{-1/2} (-1)^{s_0} $, and the fact that 
the zero-mode operator picks up $(-1)^{s_0}$ when it goes through 
the left states.\foot{
These follow from the fact that 
$ \bar{\Psi} = \exp(i \bar{H}/\sqrt{2})$ and 
left states with $s_0 = \pm 1$ are fermionic and that 
$U$ is essentially the CPT operator.
The first equation is in accord 
with $ U \Phi (\sigma)  U^{-1} = (-1)^{h} \Phi^*(-\sigma) $ with $ h $ the 
dimension of $\Phi$. Also, we can translate the boundary conditions for
$J$ and $G^\pm$ into those for $\varphi$ and $\psi^{PF}_1$.  We can confirm
that the action on $ e^{ i p x_j}$ is consistent with them.  
The action of $U$ is originally defined for $ \bar{G}^\pm $ and 
$ \bar{J} $, and hence 
there may be some ambiguity about the action on other operators such as
$ e^{i p \varphi_j} $ and $ e^{i p x_j} $. However, 
we are interested only 
in the relative phases appearing in the boundary conditions 
for various boundary states. Therefore, such an ambiguity, even if it exists,
can be absorbed into the definition of the complex field such as $ \bar{\Psi}$ 
as long as it is just an overall phase.  
}
In terms of modes, this means that 
\eqn\APsi{
   \lb  \Psi_r + i  \eta \, e^{+2\pi i \hat{Q}_\alpha} 
          \bar{\Psi}^*_{-r} \rb \ket{ \alpha } _A = 0 \period
}
For B boundary states, we obtain a similar expression 
with $ \bar{\Psi}_{-r} $ instead of $\bar{\Psi}^*$.

Since $ \Psi $ is a space-time vector, the phase appearing in \APsi\
has geometrical meaning. The phase coming from $s_i^\prime $ 
may be understood as sign ambiguity associated to fermions 
or the $Z_2$ symmetry. The remaining phase is in accord with the phase
in the previous section \xpbcktwo\ and represents the D-branes
wrapping around the cycles $\theta=\pi(m_1^\prime + m_2^\prime)/2 $
for A boundary states. 

To apply this method to the $(k=1)^3$ case is straightforward.
As a result, we get expressions similar to the above with 
\eqn\hatQkone{
   \hat{Q}_\alpha   =  
       (m'_1+m'_2+m'_3)/3 - (s'_1+s'_2+s'_3)/2 \period
}
This is in agreement with the open string channel argument.

Furthermore, we can discuss the boundary conditions for the
supercharges in a {\it generic} Gepner model. This is because
the supercharges are essentially the zero-modes of the spectral flow 
operators by half a unit and written as in \superch,\supchnl. 
Again they are
expressed by the free bosons associated to the $U(1)$ currents
and the analysis similar to the fermion case is possible. Note that 
the internal part of the spectral flow operator is nothing but $\Psi^{1/2}$
for the $(k=2)^2$ and $(k=1)^3$ case.
For the supercharges, the zero-mode part shifts $ \mu $ to $ \mu - \beta_0 $.
Consequently, we obtain
\eqn\bcQ{
    \lb  Q -\,  e^{\pi i Q_\alpha } 
          \bar{Q}^\dagger \rb \ket{ B }  = 0
    \comma 
}
for $A \otimes A$ boundary states in table 3, 
and a similar expression with $ \bar{Q} $ instead of $ \bar{Q}^\dagger$
for $B \otimes B$.  
This is the same as the result in section 6
without using free field realizations. 
The other cases for $ A(B) \otimes B(A) $ and the non-linearly realized
supercharges, $S,S^\dagger$, can be discussed similarly.
The phase for the supercharges is
half of the one in \APsi\ up to the contribution from $s'_0$. 
This is consistent 
with the fact that $ \Psi $ is a space-time vector while $Q$ is a space-time
spinor. 

\listrefs
\end